\documentclass[final]{jfm}

\usepackage{graphicx}
\usepackage{newtxtext}
\usepackage{newtxmath}
\usepackage{natbib}
\usepackage{hyperref}

\usepackage{caption}
\hypersetup{
    colorlinks = true,
    urlcolor   = blue,
    citecolor  = blue,
}

\newcommand{\RomanNumeralCaps}[1]
\nolinenumbers

\title{Influence of ambient temperature on cavitation bubble dynamics}

\author{
	Shaocong Pei\aff{1,2},
	A-Man Zhang\aff{1,2,4},
	\corresp{\email{zhangaman@hrbeu.edu.cn}}                     
	Chang Liu\aff{1,2}, 
	Tianyuan Zhang\aff{1,2}, 
	Rui Han\aff{3},
	\and Shuai Li\aff{1,2,4}
	\corresp{\email{lishuai@hrbeu.edu.cn}}}

\affiliation{\aff{1} College of Shipbuilding Engineering, Harbin Engineering University, Harbin 150001, China
\aff{2} National Key Laboratory of Ship Structural Safety, Harbin Engineering University, Harbin, 150001, China
\aff{3} Heilongjiang Provincial Key Laboratory of Nuclear Power System and Equipment, Harbin Engineering University, Harbin, 150001, China
\aff{4} Nanhai Institute of Harbin Engineering University, Sanya, 572024, China}

\begin{document}
\maketitle
 
\begin{abstract}

We examine how ambient temperature $T$ (23-90 $^\circ\mathrm{C}$) alters the dynamics of spark-induced cavitation bubbles across a range of discharge energies. As $T$ rises, the collapse of an isolated spherical bubble weakens monotonically, as quantified by the Rayleigh collapse factor, minimum volume, and maximum collapse velocity.  When the bubble is generated near a rigid wall, the same thermal attenuation is reflected in reduced jet speed and diminished migration. Most notably, at $T \gtrsim 70 ^\circ\text{C}$, we observe a previously unreported phenomenon: secondary cavitation nuclei appear adjacent to the primary bubble interface where the local pressure falls below the Blake threshold.  The pressure reduction is produced by the over-expansion of the primary bubble itself, not by rarefaction waves as suggested in earlier work. Coalescence between these secondary nuclei and the parent bubble seeds pronounced surface wrinkles that intensify Rayleigh–Taylor instability and promote fission, providing an additional route for collapse strength attenuation. These findings clarify the inception mechanism of high-temperature cavitation and offer physical insight into erosion mitigation in heated liquids.

\end{abstract}



\section{Introduction}
\label{sec:headings}

The temperature-dependent dynamics of cavitation bubbles has recently garnered significant research interest. This phenomenon is crucial not only in industrial applications, such as the erosion mitigation in high-temperature fluid machinery \citep{plesset1972temperature, phan2022thermodynamic}, structural vibrations caused by cavitation in spacecraft propulsion systems \citep{dular2018cavitation, wei2022cavitation}, and pulsed laser ablation techniques for tuning nanomaterial properties \citep{takada2010formation, wu2021effects}, but also plays a pivotal role in fundamental scientific studies, including sonoluminescence \citep{putterman2000sonoluminescence, brenner2002single} and bubble nucleation at high temperatures \citep{caupin2006cavitation,seddon2011surface}. Understanding how ambient temperature affects cavitation bubble dynamics is essential for addressing aforementioned complex challenges.

Among the challenges in bubble dynamics, the fundamental mechanisms and mitigation of cavitation erosion remain an intense and active research focus. Extensive investigations have characterized the shock waves \citep{supponen2017shock,reuter2022cavitation}, micro-jets \citep{supponen2016scaling,lechner2019fast}, and thermal loading \citep{dular2013thermodynamic,beig2018temperatures} generated during bubble collapse. In contrast, the role of ambient temperature in modulating cavitation bubble dynamics has received comparatively little attention. Limited erosion studies employing standardized specimens reveal a pronounced temperature dependence, with peak damage occurring at moderate water temperatures \citep{plesset1972temperature,dular2016hydrodynamic}. Although these observations establish a clear link between temperature and cavitation aggressiveness, the underlying physical mechanisms remain poorly understood, underscoring the need for a fundamental investigation at the single-bubble perspective.
		
Unlike other liquid properties, such as viscosity \citep{reese2022microscopic}, surface tension \citep{wu2021effects, han2022interaction}, compressibility \citep{keller1980bubble, wang2016local,han2024theoretical}, and gas saturation \citep{brenner2002single, preso2024effects}, the ambient temperature chiefly modifies the internal vapour pressure and governs heat and mass transfer at the bubble interface \citep{phan2022thermodynamic, zhang2024theoretical}. For example, laser-induced bubbles in water exhibit a pronounced dependence of oscillation period on ambient temperature \citep{barbaglia2004dependence}. In a more striking demonstration, a spark-induced cavitation bubble initiated in liquid nitrogen at -199$^\circ$C does not collapse violently; instead, it executes low-amplitude pulsations around an enlarged equilibrium radius while displaying intense interfacial instabilities \citep{chen2024investigation}. Numerical simulations by \cite{phan2022thermodynamic} corroborate these observations, showing that higher temperatures raise the vapour pressure, suppress condensation, and consequently attenuate the peak pressure reached at minimum bubble volume. Recently, \cite{geng2025} quantified the temperature dependence of spark-generated bubbles using the Keller model incorporating heat transfer, phase change, and compressibility. They attributed this high-temperature suppression to a dramatic drop in peak vapour condensation rate from 30 to 95 $^\circ \mathrm{C}$. While these findings establish that elevated temperature mitigates bubble collapse, a systematic characterization and a mechanistic explanation remain elusive. Crucially, how this temperature dependence manifests in near-wall configurations, where bubble dynamics are inherently nonspherical, remains an open and critical question. 
 
\begin{figure}
	\centering
	\includegraphics[scale=0.3]{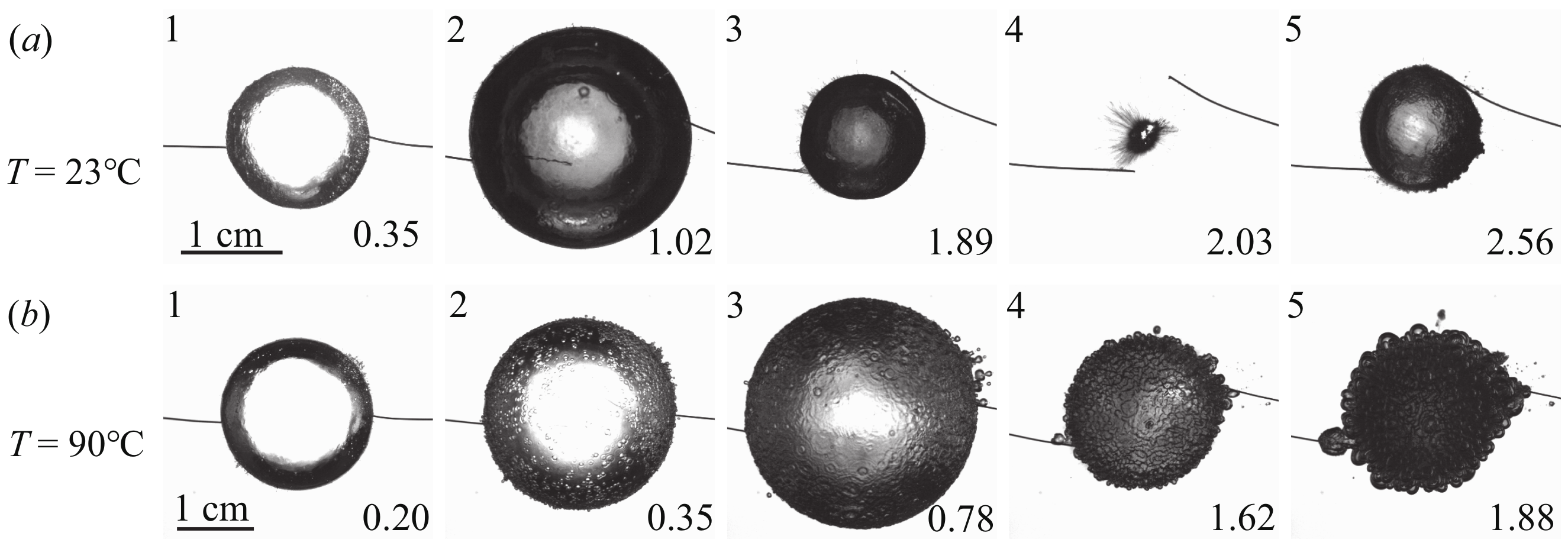}
	\caption{Representative high-speed images of spark-generated bubbles in a free field. (a) At $T= 23^\circ\mathrm{C}$, the bubble remains nearly spherical with a smooth surface during expansion, collapse and rebound. (b) At $T= 90^\circ\mathrm{C}$, the interface remains smooth during early expansion ($t=0.2$), then secondary cavitation bubbles emerge around the bubble ($t=0.35$).  These secondary cavitation bubbles coalesce into the main bubble while it is still growing ($t=0.78$), creating surface wrinkles that intensify during collapse ($t=1.62$). Upon rebound, the wrinkles develop into pronounced Rayleigh–Taylor instabilities ($t=1.88$). The time displayed in the lower right corner is normalised by characteristic collapse time, $R_{max}\sqrt{\rho/(P_{\infty}-P_v)}$, where $R_{max}$ is the maximum bubble radius, $\rho$ the liquid density, $P_\infty$ the hydrostatic pressure at infinity, and $P_v$ the saturated vapour pressure; the resulting values are 1.11 and 2.60 ms, respectively.\protect\\}
	\label{fig:0}
\end{figure}

In this study, we conducted over four hundred spark-generated cavitation bubble experiments in the free field and adjacent to a rigid wall. The extensive dataset clarifies the thermodynamics during bubble expansion, quantifies the systematic attenuation of bubble collapse, and extend the analysis to elucidate how temperature governs bubble jetting and migration near a rigid boundary. Remarkably, we identify a novel physical phenomenon in our experiments: when $T\gtrsim70^\circ \text{C}$, secondary cavitation forms near the bubble surface during the expansion phase, followed by coalescence-induced surface wrinkles. These perturbations enhance Rayleigh-Taylor instability and contribute to bubble fission, as shown in Figure \ref{fig:0}. In contrast, experiments conducted at lower temperatures do not exhibit this phenomenon. We employ the Blake's criterion to determine the origin of secondary cavitation, verify it by estimating the minimum bubble pressure, and further investigate the role of discharge energy in modulating its onset. In contrast to earlier mechanisms invoking rarefaction waves \citep{supponen2017shock, horiba2020cavitation, rossello2023bubble}, the phenomenon documented here is governed primarily by elevated ambient temperature and the intrinsic bubble dynamics. These findings advance the fundamental understanding of thermally mediated cavitation inception and suggest a new route for mitigating cavitation erosion.

The work is structured as follows. $\S$ \ref{sec:2} details the experimental setup and presents a concise yet robust method for estimating the minimum bubble pressure. $\S$ \ref{sec:3} examines bubble collapse patterns in a free field, supported by theoretical analysis using the Keller equation. $\S$ \ref{sec:4} identifies the mechanism of secondary cavitation and quantifies the influence of ambient temperature and discharge energy on its inception. $\S$ \ref{sec:5} turns to non-spherical bubble dynamics adjacent to a rigid wall, elucidating the fundamental processes relevant to cavitation erosion and complemented by a weakly compressible boundary integral simulation. Finally, the main conclusions are drawn in $\S$ \ref{sec:6}.

\section{Methodology}
\label{sec:2}

\subsection{Experimental setup}
\label{sec:2.1}
\begin{figure}
	\centering
	\includegraphics[scale=0.33]{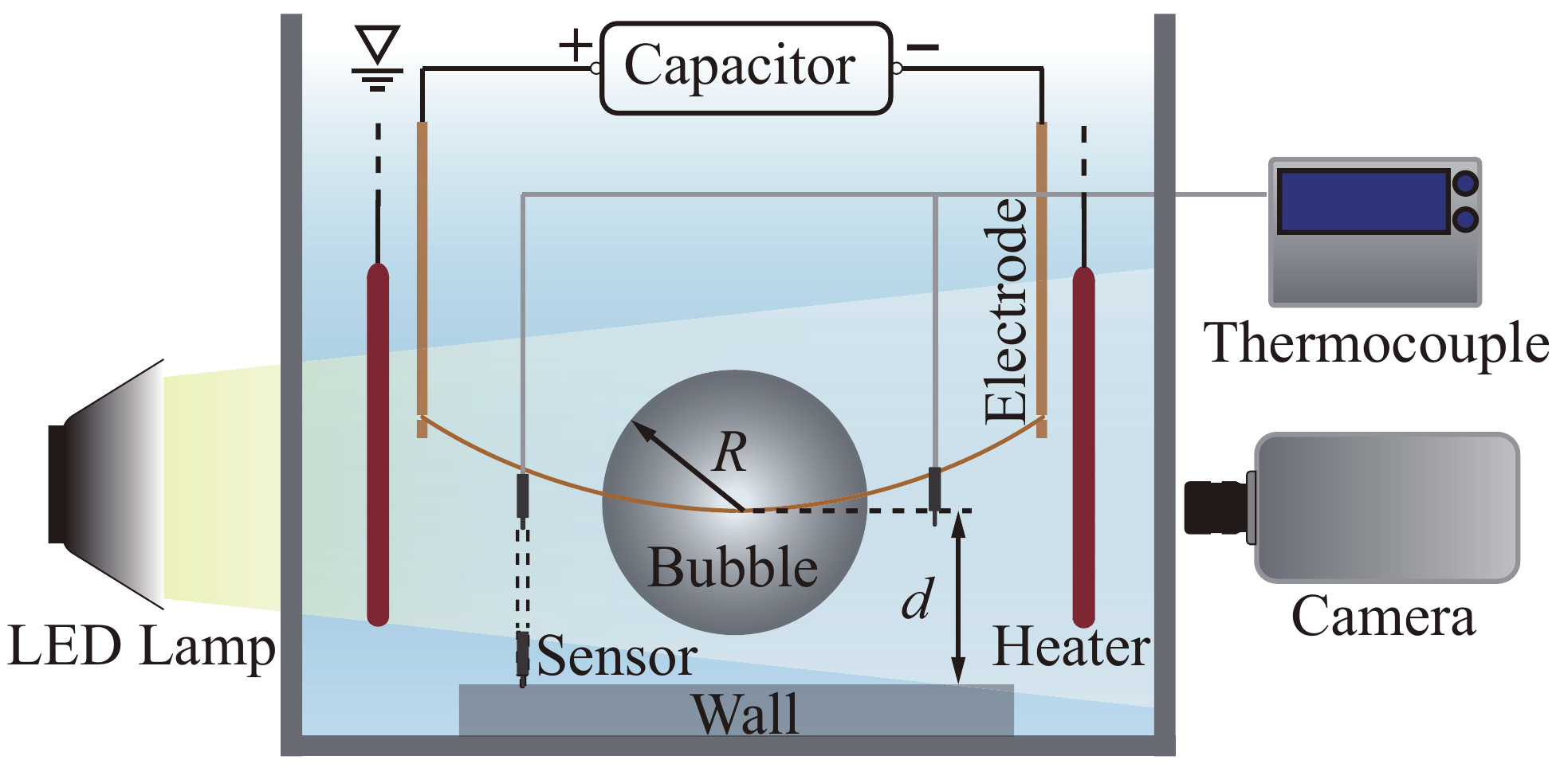}
	\caption{Experimental setup for cavitation bubble dynamics in various ambient temperatures. The bubble is generated by a underwater low-voltage electric discharge method. The temperature of water is controlled by two resistance-type heater arranged symmetrically. \protect\\}
	\label{fig:setup}
\end{figure}

In this study, cavitation bubbles were generated using an underwater low-voltage electric discharge technique \citep{turangan2006experimental,cui2018ice, han2022interaction} under controlled ambient temperature conditions. The experimental setup, illustrated schematically in Figure \ref{fig:setup}, comprises two primary subsystems: a discharge system and a heating system. A 500 $ \rm{\mu F}$ capacitor was employed as the power source. To localise bubble formation, a defect was intentionally introduced at the midpoint of the wire \citep{zhang2024theoretical}, increasing local resistance and enabling bubble generation through intense Joule heating at the defect site during discharge. To ensure statistical reliability and reproducibility, the experiment was repeated at least 15 times for each temperature.

Experiments were conducted in a transparent tempered-glass water tank with dimensions of 350 mm 
$\times$ 350 mm $\times$ 350 mm and a wall thickness of 5 mm. A PMMA plate, measuring 300 mm $\times$ 300 mm $\times$ 30 mm, was fixed at the bottom of the tank to provide a rigid wall boundary condition. The tank was filled with deionized water to a depth of 300 mm and left to stand in the test tank for at least 12 hours before experiments, ensuring it was visibly bubble-free during the tests. Two resistance-type heaters ($20 \Omega$) were used to precisely regulate the water temperature from $23$ to $90^\circ$C. Temperatures were measured using two K-type thermocouples (TASI, TA612C) positioned approximately 50 mm from the bubble centre. Within a 100 mm radius surrounding the bubble, spatial temperature variations were kept to a maximum of $0.5^\circ$C. Atmospheric pressure was monitored using a digital pressure gauge (LANGFAN, XY-3041). To capture the transient behaviours of the cavitation bubble, a high-speed camera (Phantom, V2012) equipped with a macro lens (LAOWA 100 mm F2.8) is triggered simultaneously with the discharge. Illumination is provided by a 2000W strobe-free LED lamp, which is diffused through matt glass to achieve uniform backlighting. The camera operates at $45,000-67,000$ frames per second with the exposure time of 1 $\mathrm{\mu}$s. The bubble radius $R(t)$ in our experiment is determined using the volume equivalence method, expressed as $R(t) = \sqrt[3]{3 V_b(t)/ 4\pi}$ \citep{zeng2018wall}, where $V_b(t)$ represents the time-dependent bubble volume and is obtained by integrating the bubble profile. The uncertainty in the length measurement corresponds to a single pixel, which is about  0.07 mm (approximately 0.5$\%$ of $R_{max}$).

\subsection{Pressure estimation within cavitation bubbles}
\label{sec:2.2}

We estimate the bubble internal minimum pressure at the point of bubble maximum expansion, $P_{min}$, from bubble motion equation, together with the experimentally measured velocity and acceleration of the bubble surface. Specifically, the Keller equation \citep{keller1980bubble} describes the dynamic behaviour of an initially spherical bubble oscillating in an unbounded field
\begin{equation}
	\left(1-\frac{\dot{R}}{c_\infty}\right)R\ddot{R}+\frac{3}{2}\left(1-\frac{\dot{R}}{3 c_\infty}\right)\dot{R}^2 =\left(1+\frac{\dot{R}}{c_\infty}\right)\frac{P_{l}-P_\infty}{\rho}+\frac{\dot{P_l}R}{\rho c_\infty},
	\label{2.1}
\end{equation}
where $R$ denotes the bubble radius, $c_\infty$ the speed of sound, $\rho$ the liquid density and $P_\infty$ the hydrostatic pressure at infinity. The liquid pressure at the bubble surface $P_l$, is given by
\begin{equation}
	P_l=P_b-\frac{2\sigma}{R}-\frac{4 \mu \dot{R}}{R},
	\label{2.2}
\end{equation}
with $\sigma$ and $\mu$ denoting the surface tension and dynamic viscosity, respectively. Several simplifications can be made. When estimating the minimum pressure corresponding to the moment of maximum expansion, the Mach number $\dot{R}/c$ is negligibly small in the time interval around that moment, allowing us to disregard the compressibility of the liquid. Viscosity can also be neglected because the shear rate vanishes. In addition, surface tension is safely ignored for the centimeter-sized spark-generated bubbles in this study \citep{han2022interaction}. Furthermore, we examine the P\'eclet numbers for heat and mass transport
\begin{equation}
	Pe_{h  } = \frac{U R_{max}}{\alpha} \sim O(10^{5}), \, 	Pe_{m} = \frac{U R_{max}}{D} \sim O(10^{7}),
	\label{2.3}
\end{equation}
where $\alpha$ and $D$ denote the thermal and mass diffusivities, respectively. The characteristic velocity is taken as $U=\sqrt{(P_\infty-P_v)/\rho}$. The large P\'eclet numbers, $Pe_h \sim O(10^5)$ and $Pe_m \sim O(10^7)$, suggest that both heat and mass diffusion are negligible during the bubble oscillation in our experimental system. By rearranging equation \eqref{2.1} and \eqref{2.2}, the minimum pressure within bubble can be described as

\begin{equation}
	P_{min}=P_\infty +\rho (R\ddot{R})|_{R=R_{max}}.
	\label{2.4}
\end{equation}

The maximum bubble radius, $R_{max}$, can be directly measured from experiments, while the acceleration term, $\ddot{R}$, can be derived from the temporal evolution of the bubble radius. To achieve this, we fit the bubble radius over a time interval encompassing the maximum radius using a cubic polynomial. By calculating the second derivative of this polynomial with respect to time, one can obtain $\ddot{R}$. Similar method has been applied for building thermal models of one dimensional cavitation bubbles in microtubes \citep{sun2009growth}. The chosen time interval for radius fitting is 0.25 times the first period of the bubble. This interval strikes a balance between incorporating enough experimental data for a reliable fit and keeping the period short to enhance accuracy. More details about the uncertainty method can be found in Appendix \ref{appendixA}.

\section{Spherical bubble dynamics }
\label{sec:3}

In this section, we begin by presenting the general physical phenomena of spark-generated cavitation bubbles in a free field under varying ambient temperatures. We then employ a pressure-temperature phase diagram to interpret the temperature-dependent variations in maximum bubble radius. To broaden the applicability of our conclusions, we use the maximum bubble radius $R_{max}$, the difference between the hydrostatic pressure and the saturated vapour pressure $P_\infty-P_v$, and the liquid density $\rho$ as basic quantities to convert other parameters into non-dimensional quantities. Finally, within this non-dimensional framework, we provide a quantitative analysis of key bubble collapse behaviours, including maximum bubble radius, bubble collapse time and maximum bubble collapse velocity.

\subsection{Experimental observations}
\label{sec:3.1}

\begin{figure}
	\centering
	\includegraphics[scale=0.22]{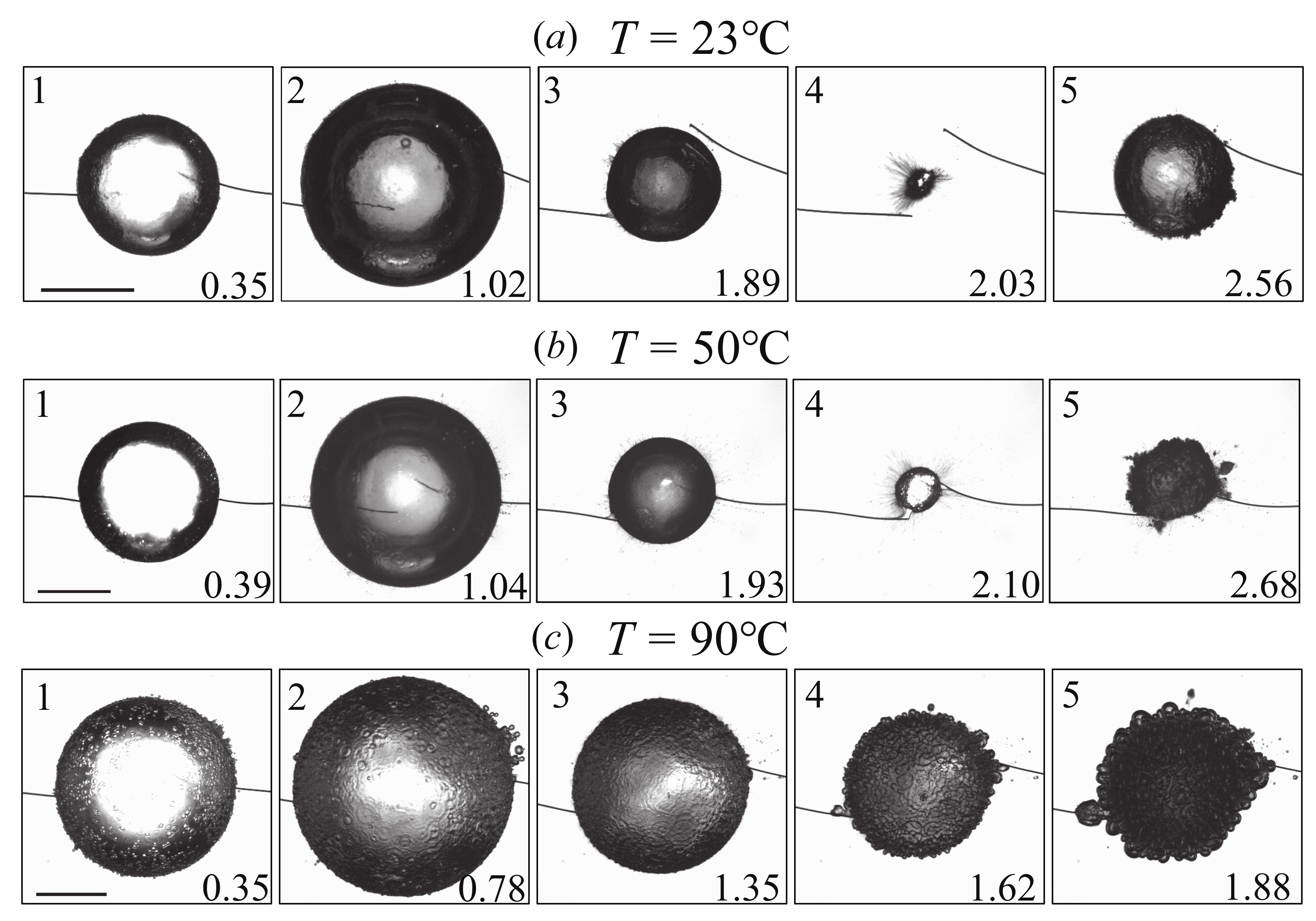}
	\caption{ Three representative experiments conducted in the free field at different ambient temperatures. (a) The bubble was initiated at room temperature (23 $^\circ \text{C}$), maintaining a spherical shape throughout the oscillation process. (b) The bubble was initiated at 50 $^\circ \text{C}$. Its contraction intensity is weakened. (c) The bubble was initiated at 90$^\circ \text{C}$. During the expansion stage, secondary cavitation occurs near the bubble wall. During the collapse stage, the intensity of bubble contraction is further weakened, and fission occurs in the rebound process. Normalised times are indicated in the lower-right corners of each frame. The time scales for non-dimensionalization  ($R_{max}\sqrt{\rho/(P_{\infty}-P_v)}  $) are 1.11, 1.38 and 2.60 ms, respectively. $P_v$ is the saturated vapour pressure corresponding to the ambient temperature. The length of horizontal line indicated in the lower-left corner of first frame is 10 mm. \protect\\}
	\label{fig:frameforfree}
\end{figure}

We conducted extensive experiments under the ambient temperature range of 23 to 90 $^\circ \mathrm{C}$ in the free field. The results from three representative experiments, each conducted at different ambient temperatures $T$, are illustrated in Figures \ref{fig:frameforfree}(a–c). Figure \ref{fig:frameforfree}(a) shows the expansion (frame 1-2), collapse (frame 3-4), and rebound (frame 5) of a bubble at room temperature ($T = 23^\circ \text{C}$). The bubble maintains nearly spherical oscillations, with a maximum radius of 11.4 mm, collapsing to a non-dimensional minimum radius $R_{min}/R_{max}$ of about 0.14. Using the method described in $\S$ \ref{sec:2.2}, we determined the minimum pressure inside the bubble (at the non-dimensional time $t \approx 1.02$) to be approximately 11,000 Pa. 
 
Figure \ref{fig:frameforfree}(b) illustrates representative moments from the experiment conducted at $T = 50^\circ \text{C}$, and the corresponding saturated vapour pressure is about 12,000 Pa. Compared to the room temperature experiment under the same discharge voltage (Figure \ref{fig:frameforfree}(a)), the maximum bubble radius $R_{max}$ increases to approximately 13.1 mm (frame 2). Subsequently, the bubble undergoes a spherical collapse with lower intensity, retaining a normalised minimum radius $R_{min}/R_{max}$ of about 0.19 at the end of the first collapse phase (frame 4). Notably, during the rebound phase, the bubble deviates from a spherical shape due to the influence of Rayleigh-Taylor instability \citep{frost1986effects, brennen2002fission}, which results in a rough and opaque surface of the bubble (frame 5). The estimated minimum pressure inside the bubble is approximately 21,000 Pa, which is about twice that in the first experiment. 

Figure \ref{fig:frameforfree}(c) shows representative bubble behaviours observed in the experiment conducted at \( T = 90^\circ \text{C} \). Compared with the previous two cases, the bubble expands to a larger maximum radius of 15.5 mm (frame 2). Additionally, the collapse of the bubble is further weakened: it reaches its minimum volume earlier, and the normalised minimum radius \( R_{min}/R_{max} \) is as large as 0.65 (frame 4). An interesting phenomenon occurs during the expansion phase: microbubbles, approximately 0.1 mm in diameter, nucleate explosively near the bubble surface (frame 2), forming a series of secondary cavitation bubbles. These secondary bubbles then migrate toward and merge with the primary bubble, creating wrinkles (pits) on its surface. These wrinkles interact and develop into chaotic patterns during the collapse phase. Driven by Rayleigh-Taylor instability, the wrinkles evolve into finger-like structures, further intensifying the fission process of the bubble (frame 5).

In the experiments conducted in this work, we find that the aforementioned secondary cavitation occurs at higher ambient temperatures, near the moment when the main bubble reaches its maximum volume in the first oscillation cycle. This phenomenon is different from the secondary cavitation caused by rarefaction waves \citep{avila2016fragmentation,supponen2017shock,rossello2023bubble}, which is independent of temperature and typically occurs after the initial expansion or collapse. In the third experiment, the estimated minimum pressure inside the bubble is about 47,000 Pa, much higher than in the previous two cases but significantly lower than the saturated vapour pressure (70,100 Pa) at this temperature. A detailed parametric discussion of this phenomenon will be provided in $\S$ \ref{sec:4}. 

		\subsection{Maximum bubble radius}
		\label{sec:3.2}
\begin{figure}
	\centering
	\includegraphics[scale=0.28]{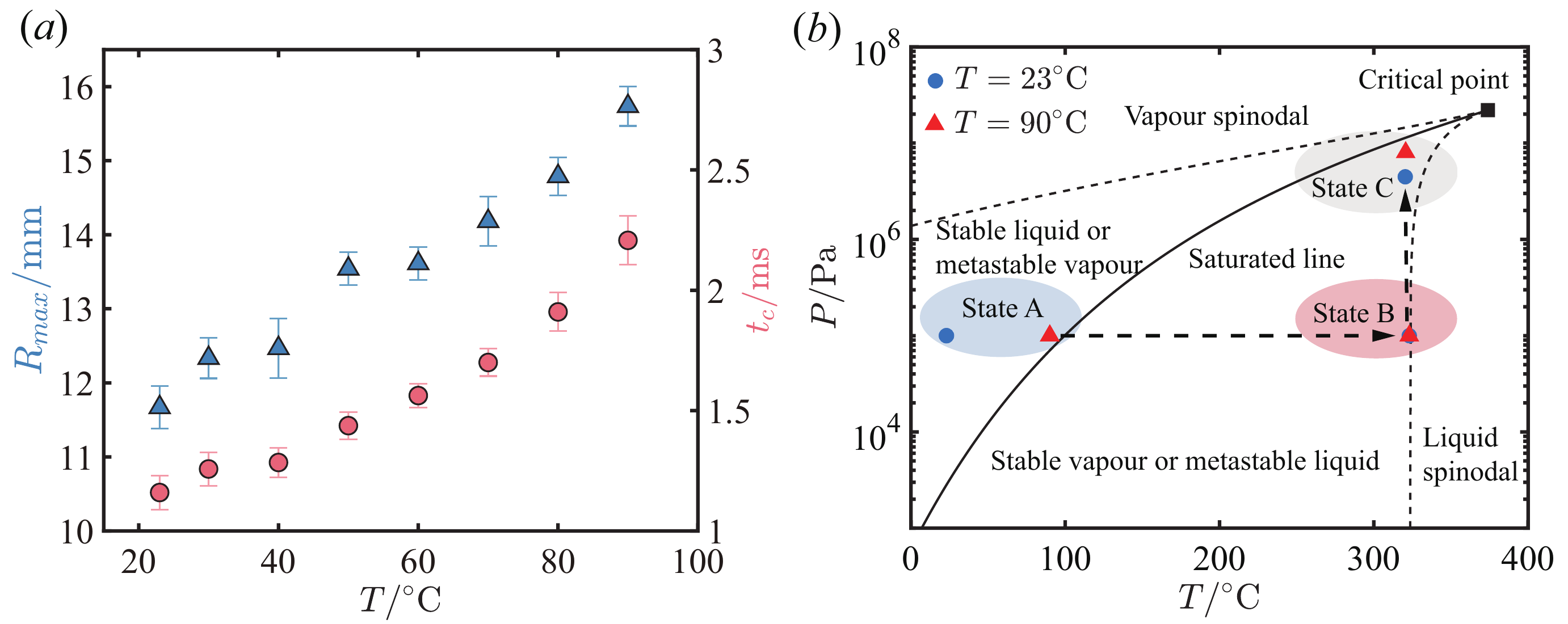}
	\caption{Effect of ambient temperature on spark-generated bubbles. (a) Variation of the maximum bubble radius $R_{max}$ (blue triangles) and collapse time $t_c$ (red circles) as functions of ambient temperature $T$. Error bars indicate the standard deviation. (b) Pressure--temperature phase diagram for spark-generated bubbles. The dashed lines denote the spinodal, representing the thermodynamic limit of stability for the vapour and liquid phases. The solid line indicates the saturated vapour pressure. Blue and red symbols mark experiments conducted at 23 and 90 $^\circ$C, respectively. State A: liquid at local hydrostatic pressure and ambient temperature before discharge. State B: liquid after energy deposition, at the spinodal limit but before vaporisation. State C: the high temperature and high pressure vapour inside the initiated bubble, corresponds to the early expansion moment when the bubble reaches the same volume in both experiments.\protect\\}
	\label{fig:Rmax}
\end{figure}

		In this section, we investigate the effect of ambient temperature on the bubble maximum radius, $R_{max}$, from a thermodynamic perspective. Figure \ref{fig:Rmax}(a) shows that both $R_{max}$ and the collapse time, $t_c$, increase clearly with temperature. This well-documented phenomenon has been reported in previous studies \citep{barbaglia2004dependence,takada2010formation,liu2013experimental}. Here, we provide a more fundamental interpretation using a pressure--temperature phase diagram to map the bubble's thermodynamic state, as shown in Figure \ref{fig:Rmax}(b). The solid line denotes the saturated vapour pressure, and the dashed lines represent the spinodal, which marks the boundary of metastable fluid states \citep{kiselev1999kinetic}. Here, it is calculated using the Peng-Robinson equation of state \citep{peng1976new}. Details are provided in Appendix \ref{appendixB}. When a liquid crosses the saturated line, it enters a metastable state where phase change is initiated by nucleation; when it continues to cross the spinodal line, the liquid becomes unstable and spontaneous phase separation occurs via spinodal decomposition. Before the electrical discharge, the state of water corresponds to State A, with both 23 and 90 $^\circ \mathrm{C}$ experiments at their respective local hydrostatic pressures. Under transient energy deposition from an electric spark, the liquid is rapidly superheated and vaporised. We infer that the phase transition mechanism involves the liquid temperature rising to the spinodal line, denoted as state B, where the liquid becomes mechanically unstable; this state is identical for both high and low temperature experiments. Our interpretation is justified by the observed melting of the copper wire. This implies that the local temperature exceeded the melting point of copper (approximately 1083$^\circ$C), well above the spinodal temperature of water (about 323$^\circ$C at 1 atm). This view of the bubble-formation mechanism has also been mentioned in previous studies \citep{vogel2005mechanisms,mohammadzadeh2017synthetic,podbevvsek2021experimental}. We nevertheless acknowledge that direct measurements of the transient temperature and pressure in the liquid, currently very challenging, would be required for definitive confirmation. As the phase transition progresses, the  rapid vaporization is confined by the inertia of the surrounding liquid, resulting in a high-temperature, high-pressure vapour bubble, denoted as state C (shown schematically). During this process, the higher ambient temperatures reduce the heat required to reach the spinodal limit, enabling a larger mass of liquid to undergo phase transition. Here, State C corresponds to the early expansion moment when the bubble reaches the same volume in both 23 and 90 $^\circ \mathrm{C}$ experiments. Since more liquid is vaporized at the higher temperature, the fixed volume at State C contains a greater number of vapour moles, leading to a higher pressure and consequently a larger bubble, as shown in Figure \ref{fig:Rmax}(a).

\subsection{Bubble collapse time}
\label{sec:3.3}

\begin{figure}
	\centering
	\includegraphics[scale=0.3]{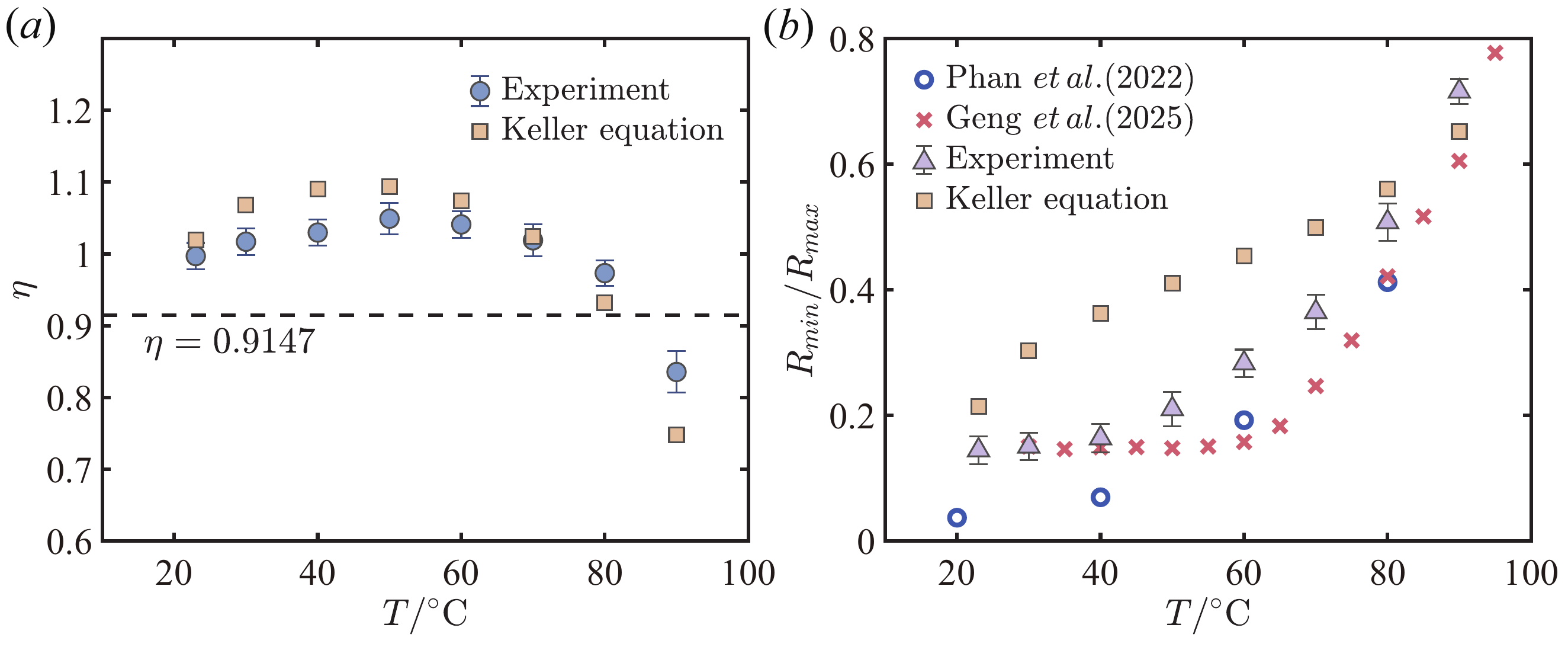}
	\caption{Effect of ambient temperature on bubble collapse. (a) Variation of the Rayleigh factor $\eta$ versus the ambient temperature $T$. The cycles denote the experimental data. (b) Variation of the non-dimensional minimum volume at the end of the first bubble collapse stage $R_{min}/R_{max}$ versus the dimensional ambient temperature $T$. The empty cycles represent the results of laser-induced bubbles from \cite{phan2022thermodynamic}. The red crosses represent the results of spark-generated bubbles from \cite{geng2025}.  The upward triangles denote the experimental data. The squares represent the results computed by the Keller equation with initial pressures of 11, 16.5, 20.7, 22.4, 28, 32, 38, and 48 kPa, which are determined by the method described in $\S$ \ref{sec:2.2}. Error bars indicate the standard deviation. \protect\\}
	\label{fig:nfactor}
\end{figure}

We have demonstrated that temperature significantly influences bubble collapse in Figure \ref{fig:frameforfree}. Here, we introduce the Rayleigh collapse factor, $ \eta = t_c / R_{max}\sqrt{\rho/ \Delta P}$, to further quantify this effect, as depicted in Figure \ref{fig:nfactor}(a). The dashed line $\eta = 0.9147$ represents the classic Rayleigh factor \citep{rayleigh1917viii}, corresponding to the collapse of a vacuum bubble under ambient pressure $P_\infty$ in an infinite domain. 

As the ambient temperature increases from 23 to 50 $^\circ \text{C}$, the average Rayleigh factor initially rises from approximately 0.99 to 1.05, exceeding the theoretical value of 0.9147. As the temperature continues to rise, this trend reverses, and $\eta$ declines, dropping to a value of 0.83 at  $T= 90^\circ \text{C}$. This reflects a continuous competition between two effects across the entire temperature range, both primarily governed by the increase in internal pressure with ambient temperature. On one hand, the elevated Rayleigh factor is primarily attributed to enhanced internal resistance during collapse, caused by compression of the bubble contents. This resistance slows down the collapse and prolongs the collapse duration. Two factors contribute to this. First, higher ambient temperature increases the vapour pressure inside the bubble. Second, when the condensation rate cannot keep up with the bubble collapse velocity, some of the vapour remaining inside the bubble behaves like a compressible gas \citep{fujikawa1980effects}. On the other hand, the minimum pressure of the bubble at its maximum size increases with the ambient temperature. The bubble therefore reaches the equilibrium radius earlier during the collapse, where internal and external pressures balance \citep{vogel1996shock}, and the collapse ends at a larger bubble radius (as shown in Figure \ref{fig:nfactor}(b)), reducing the Rayleigh factor. Note that the bubble does not collapse to a singular point, in contrast to the classical Rayleigh description. We model this process using the Keller equation,  starting from the moment of maximum expansion. The minimum bubble pressure $P_{min}$, estimated by Eq. \ref{2.4}, and the experimentally measured maximum radius $R_{max}$ are used as initial conditions. Following many previous studies \citep{wang2014multi,liang2022comprehensive,rossello2023bubble}, here we assume the total internal pressure to follow an adiabatic equation

\begin{equation}
	P_b=P_{min}\left(\frac{R_{max}}{R}\right)^{3\kappa},
	\label{eq:3.1}
\end{equation}
where $\kappa$ is the polytropic exponent. Since the content of the bubble is vapour and a small amount of non-condensable gas, we set $\kappa = 1.25$ \citep{lee2007boundary,han2022interaction}. The overall trend of the predictions agrees well with the experimental results with a maximum deviation of about $10.5 \%$ at $T=90 ^\circ$C.

We also compare the normalised minimum bubble radii in Figure \ref{fig:nfactor}(b). As can be seen, our experimental results exhibit a similar trend to those of laser-induced bubbles from \cite{phan2022thermodynamic} and spark-generated bubbles from \cite{geng2025}. The theoretical model captures the main features in the experiments, however, we notice a more pronounced difference between experiment and calculation, especially at $T \lesssim 70^\circ$C. We speculate that this discrepancy is due to the theoretical framework overlooking the effects of phase change, which play a critical role in accelerating bubble collapse through condensation at the end of the collapse. Therefore, caution is warranted when applying the adiabatic approximation in cavitation bubble modeling. Despite recent progress \citep{zhong2020model,phan2022thermodynamic}, a robust and generalizable phase change model applicable across a wide temperature range has yet to be established.

\subsection{Bubble collapse velocity}
\label{sec:3.4}

\begin{figure}
	\centering
	\includegraphics[scale=0.32]{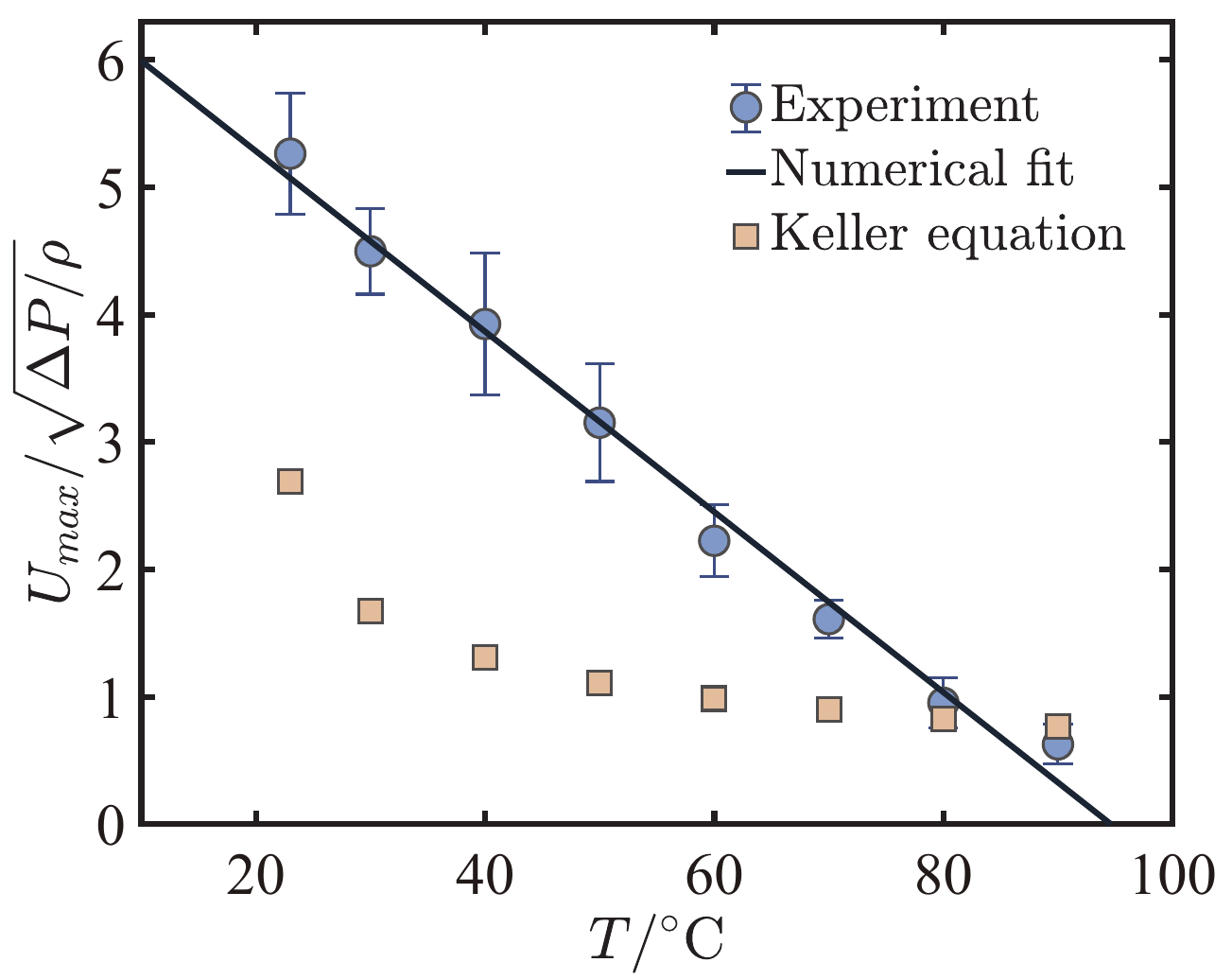}
	\caption{Variation of the non-dimensional maximum velocity during the first bubble collapse stage, $U_{max}/\sqrt{\Delta P/ \rho} $, as a function of ambient temperature  $T$.  The circles represents the experimental data and the solid line is the corresponding fitting line with a slope of -0.07 and an intercept of 6.70. The squares represent the results computed by Keller equation using the same setup as in $\S$ \ref{sec:3.3}. Error bars indicate the standard deviation. \protect\\}
	\label{fig:Umax}
\end{figure}

In this section, we examine the effect of ambient temperature on the bubble collapse velocity, providing a more direct perspective on the intensity of bubble collapse. Here we focus on the maximum collapse velocity, $U_{\text{max}} $, which is obtained by computing the frame-to-frame time derivative of the bubble radius over the entire collapse phase and taking the maximum. The normalised maximum collapse velocity $U_{max}/ \sqrt{\Delta P/\rho}$ is presented in Figure \ref{fig:Umax} as a function of the ambient temperature $T$. Due to the limited spatiotemporal resolution of the experiments, the measured velocity should be considered a lower bound of the true collapse velocity.

The non-dimensional maximum collapse velocity decreases monotonically as the ambient temperature rises from 23 to 90 $^\circ \text{C}$, dropping from approximately 5.5 to 1. The experimental data can be well fitted by a linear approximation with a slope of -0.07 and an intercept of 6.70. The root-mean-square error (RMSE) of this fit is 0.188. Figure \ref{fig:Umax} also shows the collapse velocity derived from the Keller equation, using the same setup as in $\S$ \ref{sec:3.3}. Although the theoretical results follow the same trend as our experimental observations, the predicted magnitudes are significantly underestimated, especially at temperatures below 70 $ ^\circ \text{C}$. Again, we attribute this discrepancy to the absence of phase change in the model. In reality, vapour condensation lowers the pressure inside the bubble compared to the adiabatic approximation, thereby accelerating the bubble collapse to higher velocities. Therefore, the potential for cavitation erosion may be significantly underestimated if one use the adiabatic approximation in their modelling. Nevertheless, agreement between the model and experiments improves at higher ambient temperatures. This suggests that the phase change effects become progressively weaker as temperature rises, consistent with the numerical results of \cite{phan2022thermodynamic}, who demonstrated that elevated ambient temperatures suppress vapour condensation during bubble collapse. The aforementioned discussion demonstrates that increasing the ambient temperature to elevate the vapour pressure inside the bubble is an effective strategy for mitigating the intensity of bubble collapse. A similar effect can also be achieved by modifying the fluid composition. For instance, \citet{preso2024vapour} showed that increasing the ammonia mass fraction in aqueous ammonia solutions yields a comparable effect.

\section{Mechanism of secondary cavitation}
\label{sec:4}
\begin{figure}
	\centering
	\includegraphics[scale=0.5]{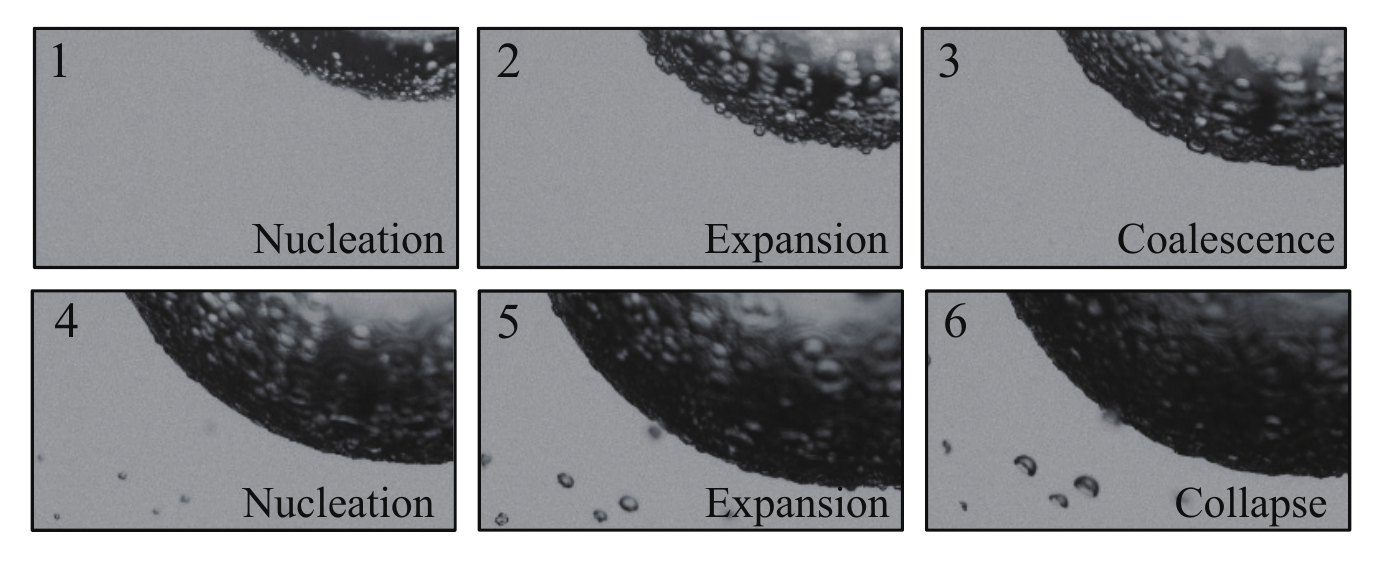}
	\caption{Experimental close-up of the evolution of secondary cavitation bubbles at 90$^\circ$C. Frames 1-3 capture the earliest instant at which secondary cavitation bubbles appear around the expanding primary bubble and subsequently coalesce with it. Frames 4-6 correspond to a later instant near maximum bubble expansion when secondary cavitation bubbles emerge at larger radial distances without noticeably perturbing the main bubble surface. The six frames occur at 0.52, 0.80, 0.99, 1.37, 2.08, and 3.08 ms. Each frame has a width of 22.3 mm.  \protect\\}
	\label{fig:closeup}
\end{figure}

  In previous sections, we observed that the Rayleigh-Taylor instability on the bubble surface becomes particularly pronounced at high temperature, leading to bubble distortion and fission during the final stages of bubble collapse. This instability arises from initial surface perturbations caused by secondary cavitation bubbles. Figure \ref{fig:closeup} presents a series of zoomed-in images illustrating the evolution of secondary cavitation bubbles in a high-temperature environment. During the expansion phase, secondary cavitation bubbles emerge around the primary bubble and subsequently coalesce with it, leading to the formation of wrinkles or pits on the gas–liquid interface (frames 1-3). This phenomenon is particularly pronounced towards the end of the expansion phase, as nucleation sites appear farther from the main interface (frames 4–6). Once secondary cavitation triggers these perturbations, the enlarged effective surface area may enhance interfacial heat transfer. The corresponding increase in baroclinic vorticity likely amplifies the surface perturbations \citep{shepherd1982rapid,frost1986effects}, manifested as progressively pronounced wrinkles on the bubble surface during the collapse phase. These surface distortions further enhance the Rayleigh-Taylor instability, promoting bubble fragmentation. \citet{chen2024investigation} also observed similar phenomena in spark-generated bubbles in liquid nitrogen.
  
\begin{figure}
	\centering
	\includegraphics[scale=0.45]{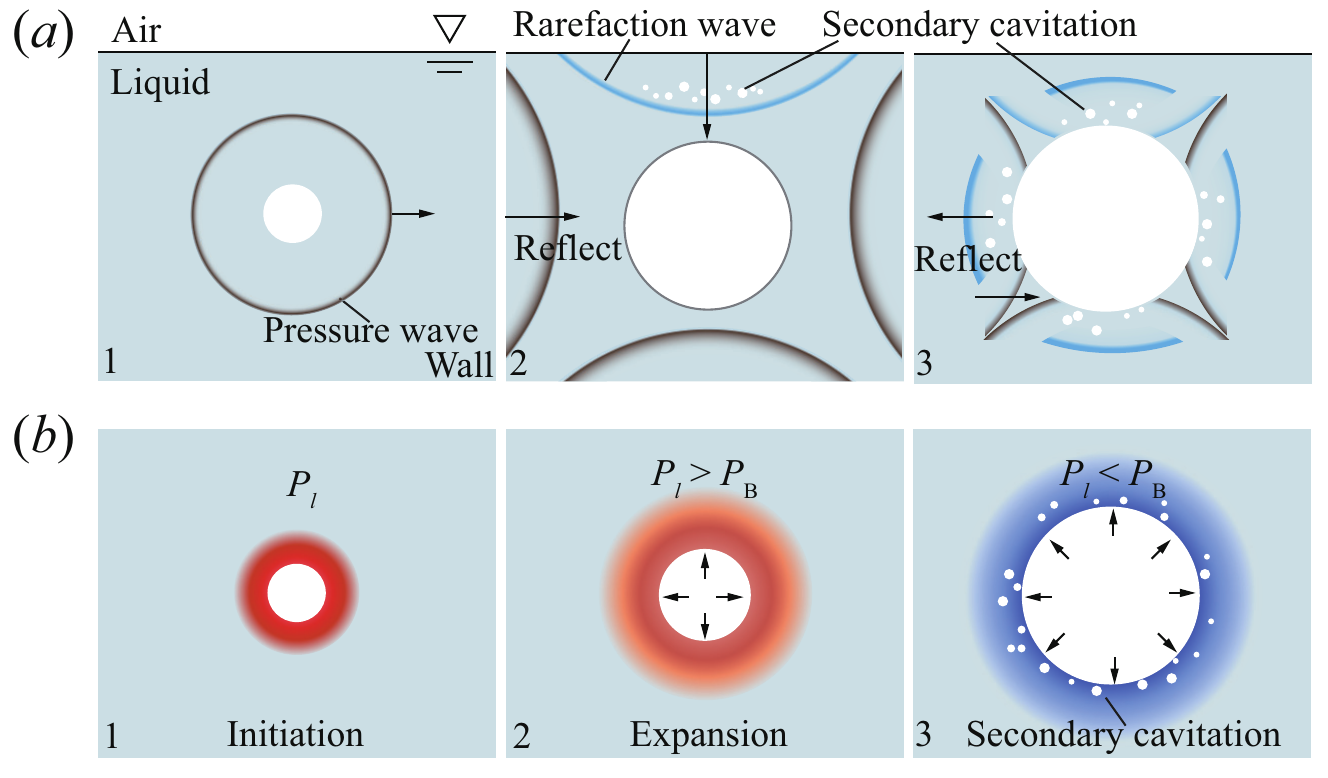}
	\caption{Schematic diagrams of two typical secondary cavitation mechanisms: (a) triggered by rarefaction waves and (b) induced by inertial expansion.\protect\\}
	\label{fig:mechanism}
\end{figure}

  The secondary cavitation phenomenon observed in this study differs from traditional secondary cavitation, which typically arises from rarefaction waves reflected at low-impedance boundaries, such as the bubble surface or free surface \citep{supponen2017shock,horiba2020cavitation, rossello2023bubble}. The spatial distribution of this cavitation is closely associated with the area influenced by rarefaction waves, as depicted in Figure \ref{fig:mechanism}(a). In our experimental setup, this mechanism can be ruled out based on the dimensions of the water tank and the timing of cavitation nucleation appearance. The secondary cavitation phenomenon observed here is characterised by its onset when the pressure at the bubble surface approaches the saturation pressure of the liquid medium at a specific temperature, as shown in Figure \ref{fig:mechanism}(b). This perspective will be elaborated below.

\begin{figure}
			\centering
			\includegraphics[scale=0.15]{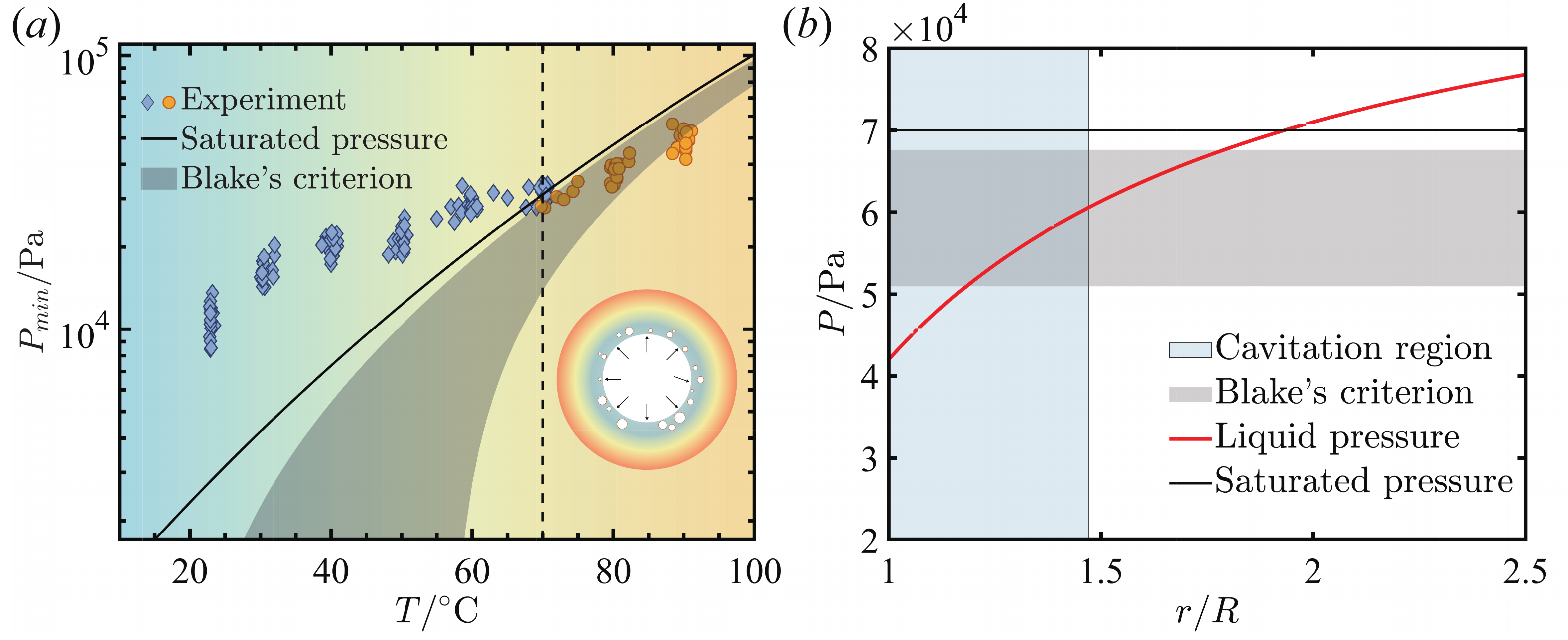}
			\caption{Mechanism of secondary cavitation. (a) Variation of the minimum pressure within the bubbles over the ambient temperature $T$. The circles denote cases where secondary cavitation occurs, while the diamonds indicate cases without secondary cavitation. (b) Radial distribution of liquid pressure at the moment of maximum expansion at 90$^\circ$C. The blue area indicates the cavitation region observed experimentally. The grey area indicates the Blake threshold pressure range corresponding to a microbubble radius of 2--10 $\rm{\mu}m$, calculated using a surface tension of 0.061 N/m. The red solid line shows the liquid pressure $P(r,t)$ as a function of $r/R$, calculated from Eq. \ref{eq:4.3}, and the black solid line denotes the saturated vapour pressure. All experiments were conducted at a discharge voltage of 200 V. \protect\\}
			\label{fig:Pmin}
		\end{figure}

		Figure \ref{fig:Pmin}(a) presents the minimum bubble pressure estimated by Eq. \ref{2.4} at different ambient temperatures. It is evident that for cases where secondary cavitation occurs (yellow circles), $P_{min}$ consistently falls below the saturated vapour pressure (solid line). To illustrate this behaviour, we introduce Blake's criterion \citep{blake1949onset,neppiras1951cavitation}, according to which cavitation inception is expected to occur when the liquid pressure falls below a critical threshold $P_B$. Under isothermal conditions, $P_B$ is given by
		\begin{equation}
			P_B= P_v-\frac{4\sigma}{3 R_B},
			\label{eq:4.1}
		\end{equation}
		where $R_B$ is the corresponding critical bubble radius
		\begin{equation}
			R_B=R_0 \sqrt{\frac{3 R_0}{2\sigma}\left(P_0+\frac{2\sigma}{R_0}\right)}.
			\label{eq:4.2}
		\end{equation}
		Here the subscript 0 denotes the equilibrium values for a microbubble in water. $R_B$ is the maximum radius for which the bubble can remain in a stable equilibrium, and $P_B$ is the corresponding minimum ambient pressure required for stability. When the ambient pressure falls below $P_B$, the bubble becomes unstable and grows without bound, which can be regarded as the onset of cavitation \citep{brennen2014cavitation,rossello2023clean}. On this basis, following the Bernoulli equation presented in \cite{plesset1977bubble}, we determine the surrounding liquid pressure $P(r,t)$
		\begin{equation}
			P(r,t)=P_\infty+\frac{R(P_l-P_\infty)}{r}+\frac{\rho R \dot{R}^2}{2r}\left[1-\left(\frac{R}{r}\right)^3\right],
			\label{eq:4.3}
		\end{equation}
		where $r$ denotes the distance from the bubble centre. We plot the radial distribution of liquid pressure at the instant of maximum bubble expansion for a 90 $^\circ$C experiment, as illustrated by the solid red line in Figure \ref{fig:Pmin}(b). The saturated vapour pressure at this temperature is 70.1 kPa, and the experimentally observed cavitation zone (blue shading) corresponds to a critical pressure of 60.6 kPa. Taking a representative microbubble radius $R_0=2-10 \rm{\mu}m$ in water \citep{brennen2014cavitation} and a surface tension of 0.061 N/m, we obtain the corresponding Blake threshold $P_B$ to be 50.9 $\sim$ 67.6 kPa (grey band). This interval brackets the observed critical pressure, confirming that secondary cavitation is triggered by the pressure drop in the surrounding liquid during bubble expansion. Now, we can easily understand the mechanism of secondary cavitation. Specifically, raising the ambient temperature increases the saturated vapour pressure and reduces the surface tension, effectively increasing $P_B$, as shown by the grey band in Figure \ref{fig:Pmin}(a). The surrounding liquid pressure at higher-temperature experiment is more prone to drop below $P_B$ during the expansion phase, thereby facilitating nucleation of secondary cavitation.

		\begin{figure}
			\centering
			\includegraphics[scale=0.27]{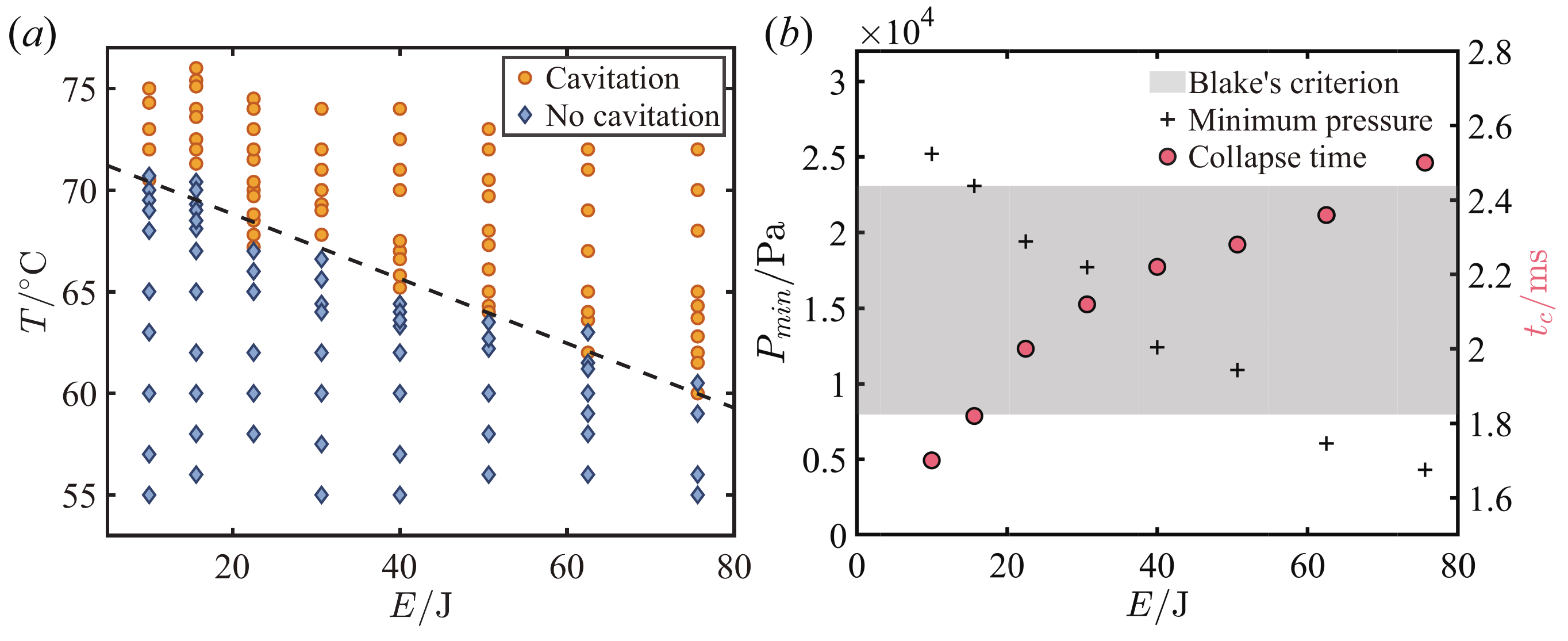}
			\caption{Influence of discharge energy on the onset of secondary cavitation. (a) Phase diagram for the occurrence of secondary cavitation with variation of the discharge energy $E$ and ambient temperature. The dashed line guides the division between regimes: circles denote cases where secondary cavitation occurs, while the diamonds indicate cases without secondary cavitation. The discharge energy $E$ is calculated as $0.5 C V_{dis}^2$ with capacitance $C=500 \mu \rm{F}$ and discharge voltage $V_{dis}$ stepped from 200 to 550 V. (b) Minimum bubble pressure $P_{min}$ (black cross, left axis) and collapse time $t_c$ (red circles, right axis) as functions of discharge energy at $T=65 ^\circ$C. The grey area indicates the Blake threshold pressure range corresponding to a microbubble radius of 2--10 $\rm{\mu}m$, calculated using a surface tension of 0.065 N/m.  \protect\\}
			\label{fig:influencefactors}
		\end{figure}

We further investigate the influence of discharge energy on the onset of secondary cavitation. Figure \ref{fig:influencefactors}(a) shows a phase diagram for the occurrence of secondary cavitation with varied discharge energy. The discharge energy $E$ is calculated as $0.5 C V_{dis}^2$ with capacitance $C=500 \mu \rm{F}$ and discharge voltage $V_{dis}$ varied from 200 to 550 V. The critical temperature of secondary cavitation decreases from approximately 70 to 60$^\circ$C as the discharge energy increases from 10 to 75.6 J. To illustrate this trend, we compare the estimated minimum bubble pressure $P_{min}$ at $T=65 ^\circ$C with the Blake threshold pressure range to be 8.0 $\sim$ 23.1 kPa, as shown in Figure \ref{fig:influencefactors}(b). As the discharge energy increases, the bubble expands more rapidly, causing a further reduction in $P_{min}$. When the discharge energy exceeds 22.5 J, $P_{min}$ falls below the Blake threshold range, providing the necessary conditions for secondary cavitation. Furthermore, we present the bubble collapse time $t_c$, which increases with discharge energy. This suggests that the low-pressure region lasts a longer time at higher discharge energies, thereby enabling small bubbles to grow to a spatially resolvable size. These two factors explain the decrease in critical temperature with increasing discharge energy and further support that secondary cavitation is triggered by the pressure drop in the surrounding liquid during bubble expansion.

\section{Bubble collapse pattern near a rigid wall}
\label{sec:5}

In this section, we systematically investigate the collapse pattern of cavitation bubbles near a rigid wall under varying ambient temperatures and standoff parameters $\gamma $, defined as the ratio of the distance $d$ between the bubble centre and the lower rigid wall to the bubble maximum radius $R_{max}$. We aim to reveal the dependence of bubble jetting, migration and the Kelvin impulse on the ambient temperature.

\subsection{Experimental observations}
\label{sec:5.1}
\subsubsection{Bubble-wall interaction at the standoff distance of $\gamma \approx$2.2}
\label{sec:5.1.1}

\begin{figure}
	\centering
	\includegraphics[scale=0.58]{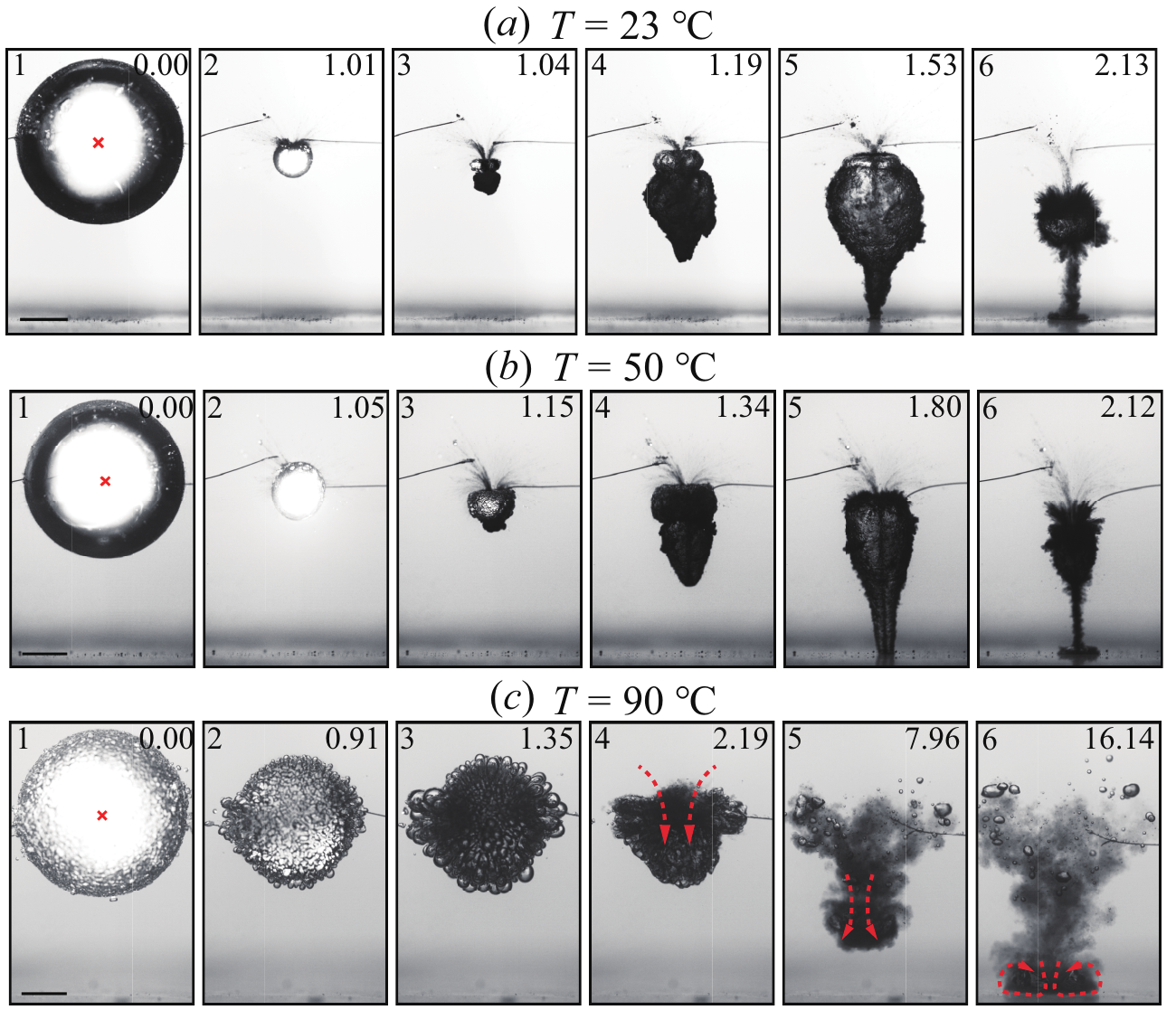}
	\caption{ Selected frames show bubbles collapsing near a solid wall at three representative ambient temperatures with a standoff parameter of $\gamma \approx 2.2$. (a) $T = 23 ^\circ \text{C}$. The jet forms and penetrates almost simultaneously at minimum bubble volume. (b) $T = 50 ^\circ \text{C}$. Both the minimum volume and oscillation period increase significantly. (c) $T = 90 ^\circ \text{C}$. The weak jet drives the bubbles toward the wall. Non-dimensional times are indicated in the top right corners of each frame. The time scales for non-dimensionalization  ($R_{max}\sqrt{\rho/(P_{\infty}-P_v)}  $) are 1.77, 1.89 and 3.26 ms, respectively. The horizontal line in the bottom left corner of the first frame is 1 cm and indicates the position of the wall. \protect\\}
	\label{fig:wall2.15}
\end{figure}

Figure \ref{fig:wall2.15} illustrates the general behaviour of bubbles at $\gamma \approx 2.2$ under varying ambient temperatures. At relatively low ambient temperatures ($T= 25$ and $50^\circ \text{C}$), the bubble expands (not shown here) and contracts almost spherically due to the weak constraint exerted by the wall, as shown in frames 1-2 of Figure \ref{fig:wall2.15}(a) and (b). However, under the secondary Bjerknes force from the wall, a jet directed toward the wall forms during the final stage of bubble collapse. As the jet travels downward, the bubble is elongated during the rebound phase (frames 4-5). During the second collapse (frame 6), the bubble is fragmented and becomes mixed with flocculent impurities, likely originating from the combustion residues of the copper alloy wire. Despite the similar general behaviours observed in the two experiments, two noteworthy observations emerge. First, the increase in temperature weakens the intensity of bubble collapse, resulting in a larger minimum radius at the end of the first cycle, consistent with the behaviour observed in a free field. Second, as the ambient temperature increases from 23 to 50 $^\circ \text{C}$, the jet velocity decreases from 185 m/s to 131 m/s.

Figure \ref{fig:wall2.15}(c) illustrates representative frames from the experiment conducted at $T=90^\circ \text{C}$. Prominent wrinkles induced by secondary cavitation appear on the bubble surface, enhancing Rayleigh-Taylor instability at the end of collapse (frames 1–2). During the rebound phase, these surface wrinkles evolve into protruding bubbles (frame 3). During the second collapse, rapid contraction of the bubble’s top surface generates a jet (frame 4). However, the jet velocity remains very low, with no significant penetration. Instead, fragmented bubbles migrate downward at approximately 1 m/s under the momentum of the jet and eventually impact the wall after a significantly long timescale (frames 5-6). As anticipated, the elevated ambient temperature substantially diminishes both the collapse intensity and jet velocity, thereby reducing the bubble's impact on the wall. 

\subsubsection{Bubble-wall interaction at the standoff distance of $\gamma \approx0.9$}
\label{sec:5.1.2}
\begin{figure}
	\centering
	\includegraphics[scale=0.52]{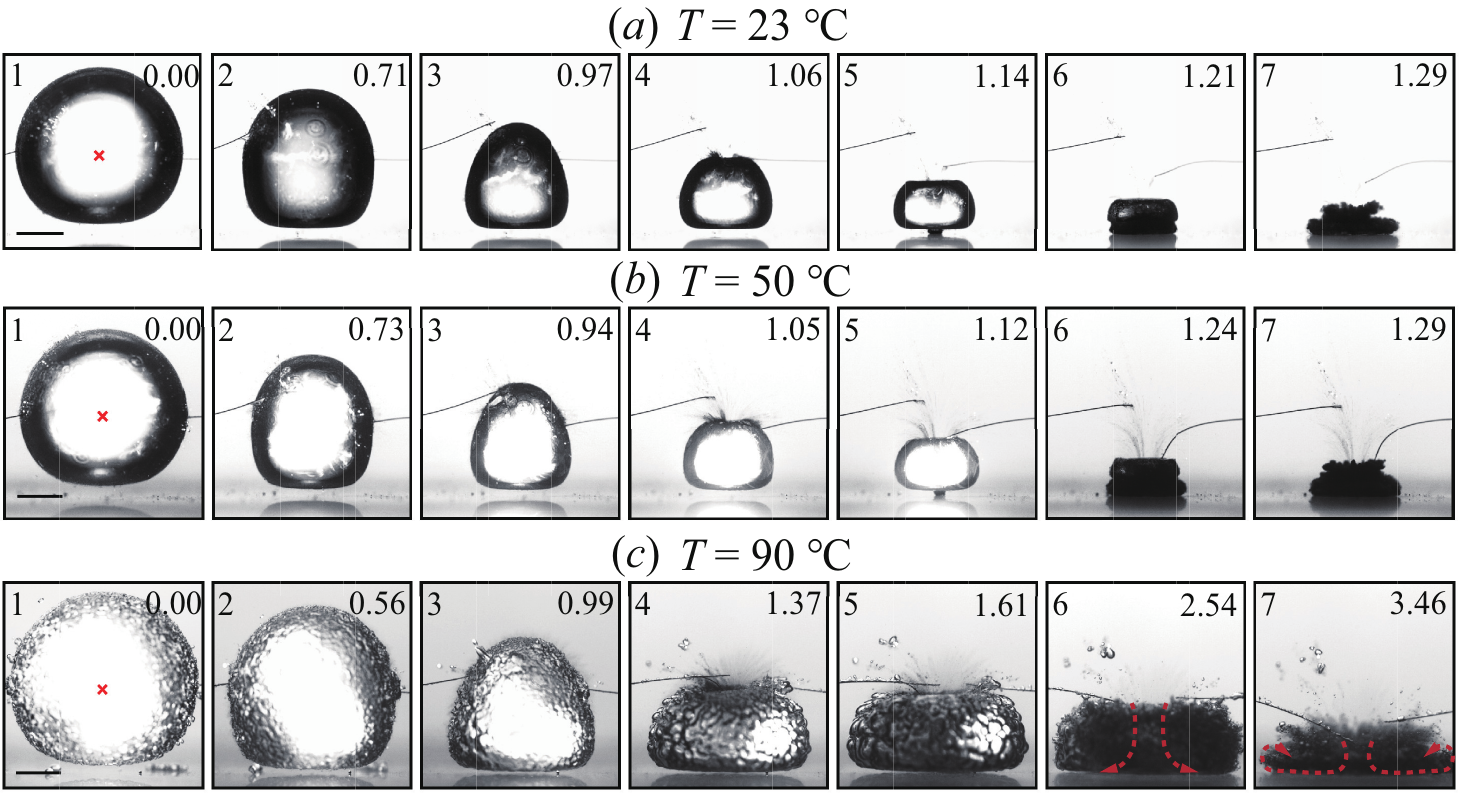}
	\caption{Selected frames show bubbles collapsing near a solid wall at three representative ambient temperatures with a standoff parameter of $\gamma \approx 0.9$. (a) $T = 23 ^\circ \text{C}$. The jet is formed early in the collapse stage and pierces the opposite wall before compressing to the bubble limit. (b) $T = 50 ^\circ \text{C}$. Bubble collapse in case (b) is similar to (a) but occurs over a longer time. (c) $T = 90 ^\circ \text{C}$. The bubble surface becomes rough under the action of secondary cavitation bubbles, similar to Figure \ref{fig:frameforfree}(c). The time and spatial scale of bubble collapse further increase. Non-dimensional times are indicated in the top right corners of each frame. The time scales for non-dimensionalization  ($R_{max}\sqrt{\rho/(P_{\infty}-P_v)}  $) are 1.78, 1.88 and 3.44 ms, respectively. The horizontal line in the bottom left corner of the first frame is 1 cm and indicates the position of the wall.  \protect\\}
	\label{fig:wall0.9}
\end{figure}

As the standoff parameter decreases, the interaction between bubble and wall strengthens. Figure \ref{fig:wall0.9} shows the bubble behaviours at $\gamma \approx 0.9$ under varying ambient temperatures. Experiments at 23 and 50 $^\circ \mathrm{C}$ exhibit highly similar bubble collapse in non-dimensional time and scale ( Figures \ref{fig:wall0.9}a, b). During collapse, liquid flow between the bubble and the lower wall is blocked, and rapid contraction in other directions drives a pronounced migration of the bubble centroid toward the wall (frames 1-3). The jet forms at an early stage of collapse and rapidly penetrates the bubble (frames 4-5). During the subsequent collapse, the bubble fully attaches to the wall (frames 6-7). Compared with the experiments at $\gamma \approx 2.2$ ($\S$\ref{sec:5.1.1}), two distinctive features emerge. First, the bubble centroid undergoes significant displacement during collapse, constrained by the wall. Second, the jets penetrate the bubble approximately 11.6$\%$ and 13.2$\%$ earlier than the final collapse, and the jet velocities in the corresponding experiments reduce to about 110 and 100 m/s, respectively.

Figure \ref{fig:wall0.9}(c) illustrates representative frames from the experiment conducted at $T=90^\circ \text{C}$. The rapid expansion of the bubble triggers secondary cavitation on both the bubble surface and the rigid wall (frames 1-3). Subsequently, a jet is formed and penetrates the bubble surface with a velocity of 23.2 m/s, occurring approximately 53.4$\%$ before the bubble reaches its minimum volume (frames 4-5). In contrast to the experiment at a larger standoff distance shown in Figure \ref{fig:wall2.15}(c), the bubble does not exhibit pronounced instability features, such as finger-like structures or disintegration. This can be explained as follows: the jet formation occupies a large amount of kinetic energy and thus the bubble is less compressed. The acceleration directed from gas phase towards the liquid phase is weakened, hindering the development of R-T instability. Upon jet impact, the jet fragments into droplets that collide with the bubble surface, inducing bubble rupture at the final collapse stage (frames 6-7).

\subsection{Jet speed}
\label{sec:5.2}

\begin{figure}
	\centering
	\includegraphics[scale=0.3]{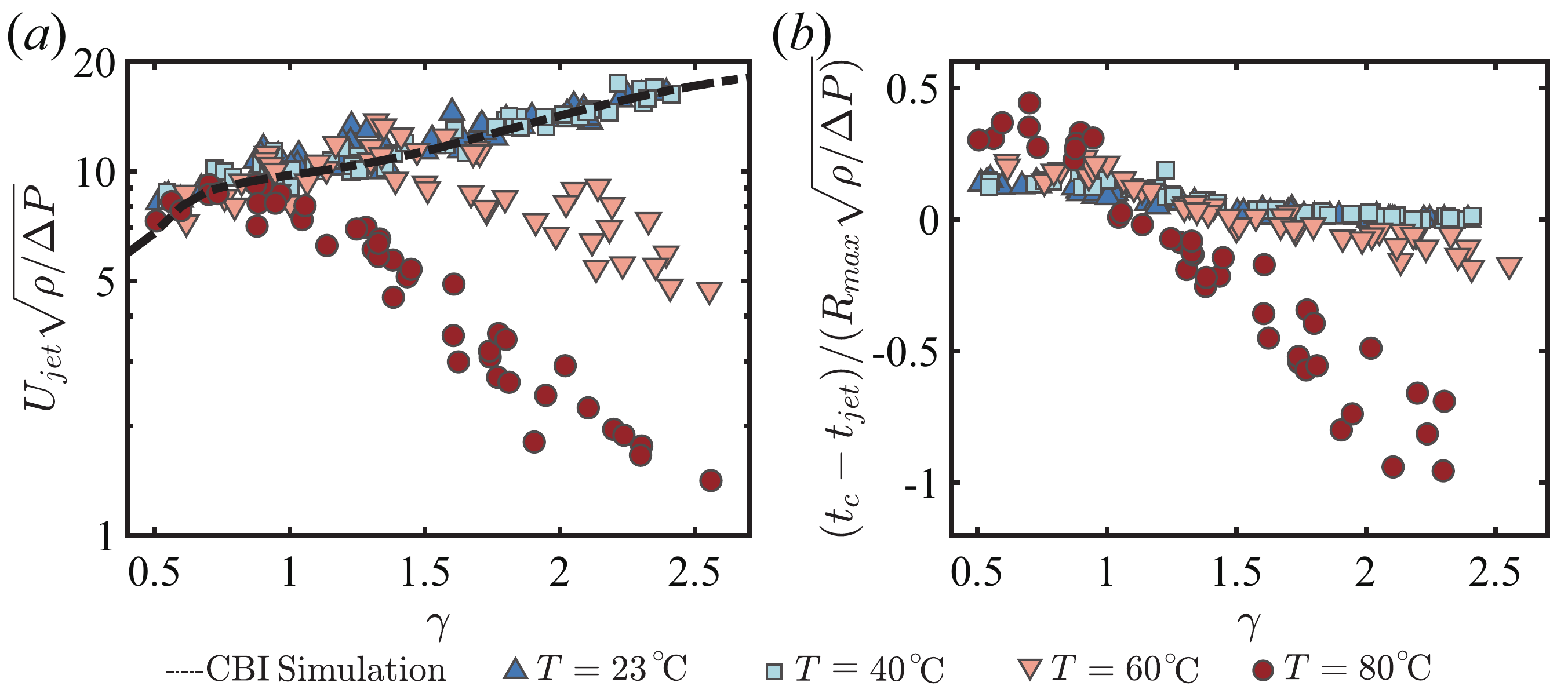}
	\caption{Dependence of jet dynamics on ambient temperature.	(a) Dependence of the non-dimensional jet velocity $U_{jet}/\sqrt{\Delta P / \rho}$ on the standoff parameter $\gamma$. (b) Dependence of the non-dimensional jet impact time $\Delta T_{jet} / R_{max} \sqrt{\rho / \Delta P}$  on the standoff parameter $\gamma$. The shapes and colors of the markers represent various ambient temperatures, as indicated in the legend. The solid line represents the results from CBI simulations. \protect\\}
	\label{fig:Ujet}
\end{figure}

In the previous section, we confirmed that the ambient temperature and standoff distance significantly influence the jet velocity. To quantify this effect, we compare the jet velocity $U_{jet}$ under four selected ambient temperatures as a function of the standoff distance ($0.5 \lesssim \gamma \lesssim 2.5$) in Figure \ref{fig:Ujet}(a). Due to the limited visibility of the bubble interior and the intense luminescence at the end of the bubble collapse, visually tracking the jet tip is highly challenging. Therefore, we determine the jet velocity by measuring the distance and time interval between its formation and the penetration point, i.e. bubble protrusion, and normalize it by the characteristic velocity $\sqrt{\Delta P / \rho}$. Additionally, we include results obtained from compressible boundary integral (CBI) simulations \citep{wang2013non,li2021comparison,li20233d,yan2025numerical}. This method is based on weakly compressible theory \citep{wang2010non} that separates the flow field into two regions: the inner flow is governed by Laplace equation, and outer flow by the linear wave equation. Matched asymptotic expansions were employed, with perturbations performed up to second order in terms of Mach number.

 Figure \ref{fig:Ujet}(a) shows experiments conducted at ambient temperatures of 23 and 40$^\circ \text{C}$. The normalised jet velocity $U_{jet}/\sqrt{\Delta P/ \rho}$ increases monotonically with the standoff parameter, rising from approximately 8.2 to 17.4. It is observed that the ambient temperature has little effect on $U_{jet}/\sqrt{\Delta P/ \rho}$. In addition, the results derived from the CBI simulation at room temperature conditions show a good agreement with the experimental data.

For experiments conducted at ambient temperatures of 60 and 80$^\circ \text{C}$, the jet velocity initially increases and then decreases with increasing $\gamma$. As $T$ rises from 60 to 80 $^\circ \text{C}$, the maximum non-dimensional jet velocity decreases from approximately 13.7 at $\gamma \approx 1.32$ to 9.2 at $\gamma \approx 0.87$. The jet velocity is strongly correlated with bubble collapse velocity at the moment of jet penetration \citep{philipp1998cavitation, supponen2016scaling}. Specifically, the closer the jet penetration occurs to the  bubble's minimum volume, the higher the resulting velocity. To illustrate the temporal relationship between jet penetration and the end of the first bubble collapse, the non-dimensional jet impact time $(t_c-t_{jet}) /( R_{max} \sqrt{\rho / \Delta P})$ is presented in Figure \ref{fig:Ujet}(b), where $t_{jet}$ denotes the time when the jet tip pierces the opposite bubble wall. At lower ambient temperatures, the jet impact time decreases and gradually approaches zero as $\gamma$ increases. This corresponds to the increase in jet velocity observed in Figure \ref{fig:Ujet}(a). At $T=60$ and $80^\circ$C, jet velocity peaks when penetration occurs slightly before the minimum bubble volume. For $T=80^\circ \text{C}$,  if $\gamma \gtrsim 1$, jet penetration occurs during rebound, resulting in a significant reduction of jet velocity (Figure \ref{fig:Ujet}(a)). 

\subsection{Bubble displacement}
\label{sec:5.3}

\begin{figure}
	\centering
	\includegraphics[scale=0.3]{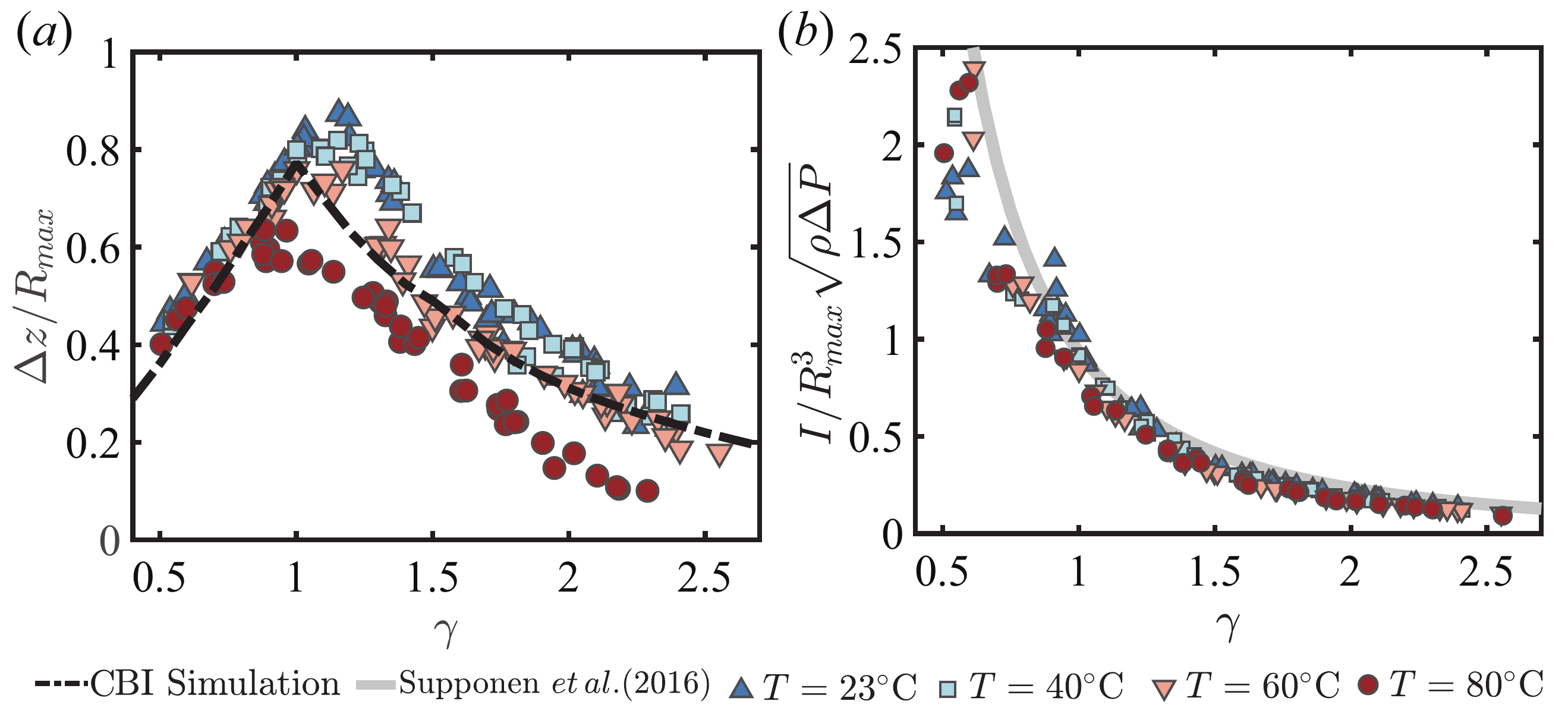}
	\caption{Bubble motion during the first cycle at different ambient temperatures. (a) Variation of the non-dimensional displacement of bubble centroid $\Delta z / R_{max}$ versus the standoff parameter $\gamma$. (b) Variation of the non-dimensional Kelvin impulse $I / R _{max}^3 \sqrt{\rho \Delta P}$ versus the standoff parameter $\gamma$. The shapes and colours of the markers represent various ambient temperatures, as indicated in the legend. The dashed lines represent the best fit for experimental data.   \protect\\}
	\label{fig:disp}
\end{figure}

In this section, we investigate the effect of ambient temperature on bubble centroid displacement, a critical parameter closely associated with the wall damage caused by jets or shock waves. Specifically, $\Delta z$ is defined as the displacement from bubble inception to the end of its first collapse. Figure \ref{fig:disp}(a) compares the non-dimensional centroid displacement, $\Delta z/ R_{max}$, across four ambient temperatures as a function of standoff parameter $\gamma$. The results obtained from CBI simulations are also included in Figure \ref{fig:disp}(a) for comprehensive analysis.

As bubble inception approaches the rigid wall, centroid displacement initially increases and then decreases, with the transition occurring at a standoff distance of approximately $\gamma =1$. Our CBI simulations effectively capture this feature under relatively low ambient temperatures. For $\gamma \gtrsim 1$, decreasing $\gamma$ enhances the anisotropy of the pressure field induced by the wall, resulting in a more pronounced tendency for the bubble to migrate towards the wall. As $\gamma$ further decreases below 1, the proximity of the boundary constraints bubble migration. Notably, under an ambient temperature of 80$^\circ \text{C}$, when the bubble is far from the boundary ($\gamma \gtrsim 1$), the centroid displacement is significantly lower than that observed in experiments conducted at lower temperatures. 
		
To further elucidate the underlying mechanism, we introduce the Kelvin impulse $I$ \citep{blake1982note}, an important concept widely used to predict the direction of bubble translation and jet formation \citep{blake2015cavitation,supponen2016scaling,kang2019gravity,han2022interaction}. The Kelvin impulse is defined as

\begin{equation}
	I=\int_{0}^{t_c}F\mathrm{d} t,
\end{equation}
where $F$ is the force acting vertically downward on the field, comprising two components
\begin{equation}
	F=F_{b}+F_{g},
\end{equation}
with $F_{b}$ and $F_{g}$ denoting the secondary Bjerknes force and buoyancy, respectively. These forces are formulated as
\begin{equation}
	F_{b}=-\frac{\rho m^2}{16 \pi d^2},
\end{equation}

\begin{equation}
	F_{g}=\rho gV_b.
\end{equation}
The source strength $m$ is calculated using the expression $4 \pi R^2 \dot{R}$, where the bubble wall velocity $\dot{R}$ is derived from the time derivative of the bubble radius. In Figure \ref{fig:disp}(b), we present the normalised Kelvin impulse $I / R^3_{max} \sqrt{\rho \Delta P}$ as a function of $\gamma$. For comparison, we also include the normalised Kelvin impulse $0.934 \gamma ^{-2}$ \citep{supponen2016scaling}. Surprisingly, the normalised Kelvin impulse is independent of ambient temperature and agrees well with theoretical results when $\gamma > 1$. It should be noted that the dimensional Kelvin impulse decreases with ambient temperature. When $\gamma < 1$, the bubble deviates significantly from a spherical shape, placing it beyond the applicability of the Kelvin impulse theory.

In the non-dimensional framework, the centroid displacement of a cavitation bubble is influenced by its added mass. As the Kelvin impulse becomes approximately constant at the end of collapse, the migration velocity scales inversely with that added mass, which itself scales with bubble volume \citep{philipp1998cavitation}. In our experiments, a higher ambient temperature leads to a larger minimum bubble volume. The associated increase in added mass reduces both the migration velocity and the resulting centroid displacement. This explains the dependence of the bubble displacement on ambient temperature.

\section{Conclusions}
\label{sec:6}

This study investigated the influence of ambient temperature $T$ (23-90$^\circ \mathrm{C}$) on the dynamics of spark-generated cavitation bubbles at discharge energies spanning 10-75.6 J. Over four hundred experiments were conducted to quantify the thermodynamics of bubble expansion, to document the systematic attenuation of bubble collapse, and to elucidate near-wall collapse patterns.  Complementary insights were provided by the Keller equation and compressible boundary integral simulations, which also helped delineate the validity range of the adiabatic approximation for the internal bubble pressure. A concise method for estimating the internal bubble pressure, in conjunction with Blake’s criterion, elucidates the mechanism of secondary cavitation and its dependence on ambient temperature and discharge energy. The main findings of this study are drawn below.

For spherical bubbles in the free field, the maximum radius increases monotonically with ambient temperature $T$, as less superheat is required to approach the spinodal limit for vaporization. However, the collapse becomes progressively weaker as $T$ increases, as evidenced by (i) the non-dimensional minimum radius $R_{\min}/R_{\max}$, (ii) the normalised peak velocity $U_{\max}/\sqrt{\Delta P/\rho}$, and (iii) the minimum internal pressure $P_{\min}$. Increasing $T$ from 23 to 90$^\circ \mathrm{C}$ raises $R_{\min}/R_{\max}$ from $0.14$ to $0.7$, reduces $U_{\max}/\sqrt{\Delta P/\rho}$ almost linearly from $5.5$   to $1$, and elevates $P_{\min}$ from $10^4$ Pa to $4.9 \times 10^4$ Pa (10 J discharge). Keller-equation solutions further reveal that the adiabatic assumption for the bubble interior pressure fails at low temperatures but remains valid at high temperatures ($\gtrsim$ 70$^\circ \mathrm{C}$ ).

Beyond the primary spark-generated bubble, we identify a previously unreported phenomenon that appears at ambient temperatures above $70 ^\circ \text{C}$ (10 J discharge): secondary cavitation nuclei emerge adjacent to the primary bubble surface and subsequently coalesce into surface wrinkles. We show that these nuclei form where the local liquid pressure drops below the Blake threshold. This pressure drop is driven by the over-expanded bubble itself, rather than by the rarefaction waves invoked in earlier studies. Quantifying the influence of discharge energy $E$ reveals that increasing $E$ from 10 to 75.6 J lowers the critical temperature for inception from 70 to $60^\circ\mathrm{C}$. Such interfacial perturbations amplify Rayleigh-Taylor instabilities and promote bubble fission during collapse.

When a bubble collapses near a rigid wall with standoff parameter $0.5<\gamma<2.5$, the erosion potential decreases as $T$ rises.  This trend is evidenced by the non-dimensional jet velocity $U_{\text{jet}}/\sqrt{\Delta P/\rho}$ and by the centroid displacement $\Delta z/R_{\max}$.  At $T=23$ and $40 ^\circ \text{C}$, $U_{\text{jet}}/\sqrt{\Delta P/\rho}$ is almost insensitive to temperature and grows monotonically with $\gamma$, consistent with earlier studies.  At $T=60$ and $80 ^\circ \text{C}$, however, $U_{\text{jet}}/\sqrt{\Delta P/\rho}$ first increases with $\gamma$, reaches a maximum at an intermediate $\gamma$ (where the jet penetrates slightly before the final collapse), and then declines. Although the dimensional Kelvin impulse decreases with $T$, the normalized Kelvin impulse collapses onto a single curve for all temperatures and obeys the classical scaling $I/R_{max}^3\sqrt{\rho/\Delta P} \sim \gamma^{-2}$ for $\gamma\gtrsim0.9$.  The added mass of the bubble increases with $T$. The larger added mass lowers the migration velocity and, consequently, the centroid displacement $\Delta z/R_{\max}$.

\backsection[Acknowledgements]{The authors thank Y. Sun, K. Lyu, T. Ding, and L.   Lu from HEU for their valuable assistance in conducting the experiments.}

\backsection[Funding]{This work is supported by the National Natural Science Foundation of China (nos. 12372239, 52525102), the Key R$\&$D Program Project of Heilongjiang Province (JD24A002), National Key Laboratory of Ship Structural Safety (Naklas2024ZZ004-J), and the Xplore Prize. }

\backsection[Declaration of interests]{The authors report no conﬂict of interest.}

\backsection[Data availability statement]{The data that support the findings of this study are openly available in [repository name] at http://doi.org/[doi], reference number [reference number]. See JFM's \href{https://www.cambridge.org/core/journals/journal-of-fluid-mechanics/information/journal-policies/research-transparency}{research transparency policy} for more information}

\backsection[Author ORCIDs]{Shaocong Pei https://orcid.org/0009-0008-7098-8137; A-Man Zhang https://orcid.org/0000-0003-1299-3049; Chang Liu https://orcid.org/0009-0008-2325-2813; Tianyuan Zhang https://orcid.org/0000-0001-7238-2608; Rui Han https://orcid.org/0000-0003-3699-5954; Shuai Li https://orcid.org/0000-0002-3043-5617;}


\appendix

		\section{Verification of pressure estimation within cavitation bubbles}\label{appendixA}
		
		\begin{figure}
			\centering
			\includegraphics[scale=0.3]{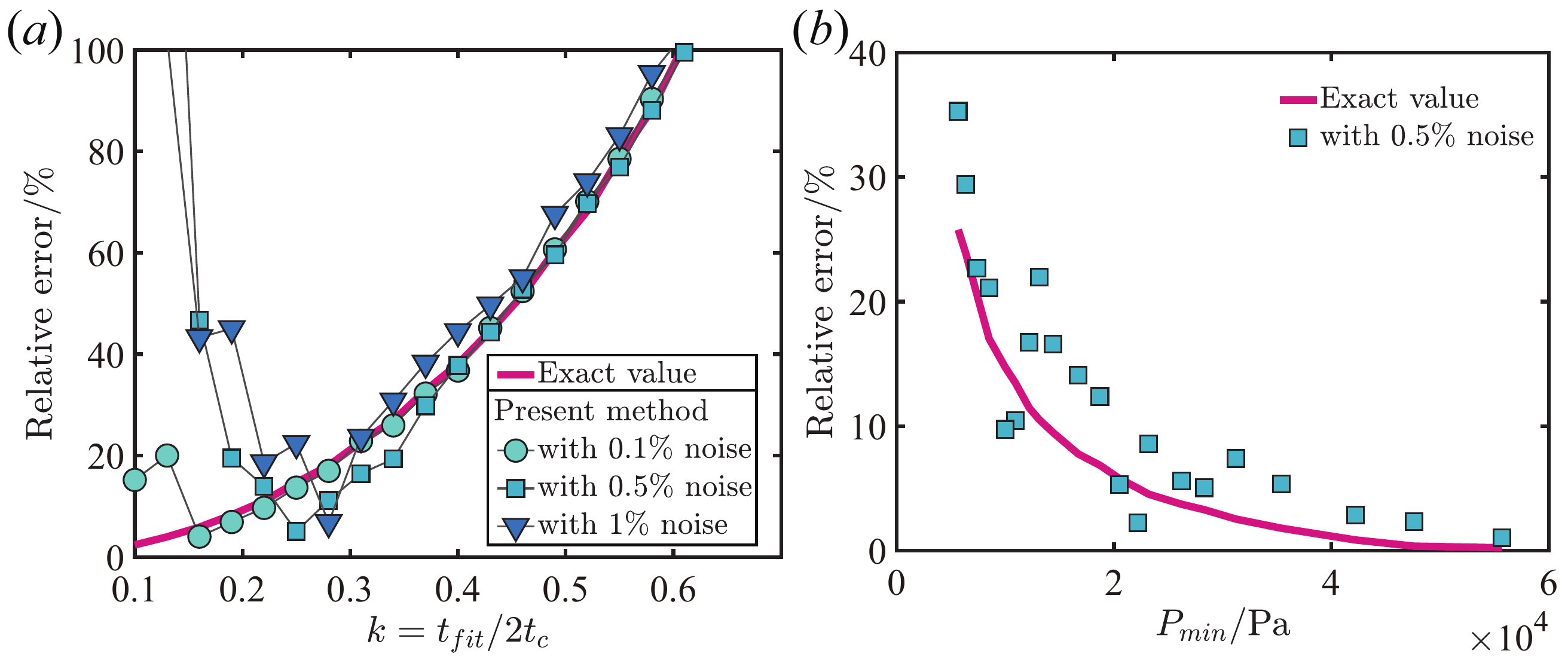}
			\caption{Accuracy assessment of the method used to estimate the pressure inside cavitation bubbles. (a) Relative error as a function of $k$. The Keller solution yields a minimum internal pressure of $1.1 \times 10^4$ Pa. The theoretical radius curve is sampled at the experimental temporal resolution, and Eq. \ref{2.4} is used to recover the minimum pressure from these synthetic data. Solid magenta lines represent noise-free data. Circles, squares, triangles denote data with noise amplitudes of 0.1\%, 0.5\%, and 1\% of $R_{max}$ on the clean radius history. (b) Relative error versus $P_{min}$.}
			\label{fig:accuracy}
		\end{figure}

		This section quantifies the accuracy of the method used to estimate the pressure inside cavitation bubbles. As described in $\S$ \ref{sec:2}, we approximate the bubble radius history over an interval that encloses the instant of maximum radius with a cubic polynomial. A key adjustable parameter is the time interval employed for the fit, expressed through the non-dimensional span ratio $k$--the selected interval scaled by the first bubble oscillation period. To assess the sensitivity to $k$, we generate reference data with the Keller equation for the following parameters: initial bubble radius $R_{0}=2.228 $mm, initial internal pressure 100 $P_\infty$, the polytropic exponent $\kappa=1.25$, speed of sound $c_\infty$=1500 m/s, liquid dynamic viscosity $\mu$ =0.001 Pa·s, and surface tension coefficient $\sigma$= 0.073 N/m. The Keller solution yields a minimum internal pressure of $1.1 \times 10^4$ Pa. The theoretical radius curve is then sampled at the experimental temporal resolution, and Eq. \ref{2.4} is used to recover the minimum pressure from these synthetic data. Figure \ref{fig:accuracy}(a) shows the relative error of the method as a function of $k$. Without added noise the error grows monotonically with $k$, because the cubic polynomial progressively fails to capture the true radius evolution. Real measurements are limited by spatial resolution. To examine this effect, we superimpose artificial noise of amplitudes 0.1$\%$, 0.5$\%$, and 1$\%$ of $R_{max}$ on the clean radius history. With noise present, the relative error first decreases with fluctuations as $k$ increases, then gradually rises and finally approaches the noise-free curve. This behaviour indicates that the fit is most vulnerable to noise at small vulnerable. Given our imaging resolution of 14 pixels per millimetre ($\approx $0.5$\%$ of $R_{max}$), we adopt $k=$ 0.25, which keeps the relative error below approximately 10$\%$. Building on this, Figure \ref{fig:accuracy}(b) further shows how the relative error scales with  $P_{min}$. The error rises sharply as $P_{min}$ decreases, because the steeper acceleration near maximum expansion amplifies the truncation error of the cubic polynomial. Across the entire pressure range examined, the relative error remains below approximately $15\%$ in the present study. It should be noted that this method is most accurate when the bubble pressure is close to the ambient hydrostatic pressure. As bubble pressure drops, the uncertainty grows. If it approaches or falls below the vapour pressure, a phase change model is needed to keep the results reliable.

		
		

		\section{Details of Spinodal Determination}\label{appendixB}
		
		The spinodal line is determined from its definition, $(\partial P / \partial V_m)_T = 0$, using the Peng-Robinson equation of state \citep{peng1976new}, given by
		\begin{equation}
			P = \frac{\mathcal{R}T}{V_m-b}-\frac{a\alpha}{V_m^2+2bV_m-b^2},
		\end{equation}
		where $\mathcal{R}$ is the universal gas constant and $V_m$ is the molar volume. Here $T$ is the absolute temperature in kelvins ($\mathrm{K}$). The parameters $a$, $b$, and $ \alpha$ represent intermolecular attraction, molecular repulsion, and temperature-dependent attraction, respectively
		\begin{gather}
			a=0.45724 \frac{\mathcal{R}^2T_{cr}^2}{P_{cr} ^2}, \\
			b=0.0778\frac{\mathcal{R}T_{cr}}{P_{cr}}, \\
			\alpha= (1+k(1-\sqrt{\frac{T}{T_{cr}}}))^2,\\
			k\approx 0.37464+1.54226\omega-0.26992\omega^2,
		\end{gather}
		with acentric factor $\omega=0.344$ for water \citep{coolprop}, and subscript $cr$ denoting the critical point.

		\section{Secondary cavitation in laser-induced bubble experiments}\label{appendixC}
\begin{figure}
			\centering
			\includegraphics[scale=0.58]{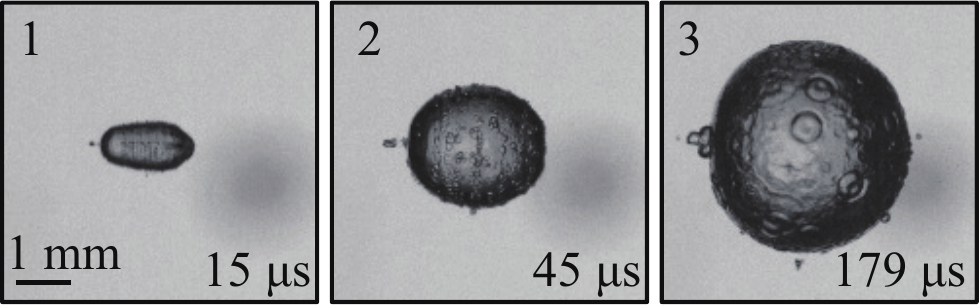}
			\caption{Laser-induced cavitation bubble in a free field at 80 $^\circ$C. The bubble, initially spindle-shaped, undergoes rapid expansion, with secondary cavities nucleating around it at approximately $45 \mu\mathrm{s}$. By $179 \mu\mathrm{s}$, the coalescence of these bubbles produces pronounced surface wrinkles.	  \protect\\}
			\label{fig:laser}
		\end{figure}

		Preliminary laser-induced cavitation experiments were conducted to test the universality of the secondary cavitation phenomenon and to exclude any influence of spark-generated debris.  The experimental setup follows \cite{li2024cavitation} and \cite{zhao2025new}.  Representative high-speed images at $80^\circ\mathrm{C}$ (Figure \ref{fig:laser}) show an initially spindle-shaped bubble that expands rapidly.  Secondary cavitation nuclei appear around the primary bubble at $t=45 \mu\mathrm{s}$. By $t=179 \mu\mathrm{s}$, when the bubble attains its maximum volume, the coalesced bubbles have imprinted pronounced wrinkle-like disturbances on the interface. This reproducible sequence confirms that residual impurities from spark discharge have a negligible effect on our conclusions.

\bibliographystyle{jfm}

\bibliography{Reference}


\end{document}